\documentclass[12pt]{article}
\usepackage{authblk}
\usepackage{hyperref}
\usepackage[letterpaper,margin=1in]{geometry}
\usepackage{graphicx}
\usepackage{lipsum, wrapfig}
\usepackage[labelfont=bf]{caption}
\usepackage{sectsty}
\usepackage{indentfirst}
\usepackage{titlesec}
\usepackage{amsmath}
\usepackage[normalem]{ulem}
\usepackage{multicol}
\usepackage{caption}
\begin{document}

\title{The physicist's guide to one of biotechnology's hottest new topics: CRISPR-Cas}

\author
       {Melia E.\ Bonomo$^{1,3}$ and Michael W.\ Deem$^{1,2,3}$
       \\
       {\small $^1$\emph{Department of Physics and Astronomy, Rice University, Houston, TX 77005, USA}}\\
      {\small $^2$\emph{Department of Bioengineering, Rice University, Houston, TX 77005, USA}}\\
       {\small $^3$\emph{Center for Theoretical Biological Physics, Rice University, Houston, TX 77005, USA}}}

\date{}
\maketitle

\tableofcontents

\begin{abstract}
Clustered regularly interspaced short palindromic repeats (CRISPR) and CRISPR-associated proteins (Cas) constitute a multi-functional, constantly evolving immune system in bacteria and archaea cells.  A heritable, molecular memory is generated of phage, plasmids, or other mobile genetic elements that attempt to attack the cell.  This memory is used to recognize and interfere with subsequent invasions from the same genetic elements.  This versatile prokaryotic tool has also been used to advance applications in biotechnology.  Here we review a large body of CRISPR-Cas research to explore themes of evolution and selection, population dynamics, horizontal gene transfer, specific and cross-reactive interactions, cost and regulation, non-immunological CRISPR functions that boost host cell robustness, as well as applicable mechanisms for efficient and specific genetic engineering.  We offer future directions that can be addressed by the physics community.  Physical understanding of the CRISPR-Cas system will advance uses in biotechnology, such as developing cell lines and animal models, cell labeling and information storage, combatting antibiotic resistance, and human therapeutics.
\end{abstract}

\section{Introduction}

In 1987, Ishino and colleagues had set out to identify the encoded protein and primary structure of a particular gene in \emph{Escherichia coli} by analyzing its chromosomal DNA segment and flanking regions~\cite{75}.  They found an interesting sequence structure at the gene's 3'-end flanking region, in which five homologous sequences of 29 nucleotides were arranged as direct repeats with 32-nucleotide sequences spaced between them. Little did they know that their discovery would prove to have critical immunological significance.  It was not until 2000 that these mysterious repeated genomic elements were revisited when Mojica and colleagues searched the available microbial genome database and found many organisms that contained partially palindromic sequences of 24--40 basepairs with 20--58 basepair sequences spaced between them~\cite{47}.  These were found in almost all archaea, about half of bacteria, no viruses, and no eukaryotes.  Related and unrelated species had nearly identical structure in these repeat sequence units.  The sequences in between, called `spacers,' were unique to an individual locus and were not found in other genomes~\cite{35}.  After many suggested abbreviations, including SRSRs, {\bf s}hort {\bf r}egularly {\bf s}paced {\bf r}epeats, and SPIDR, {\bf sp}acers {\bf i}nterspersed {\bf d}irect {\bf r}epeats, the scientific community settled on calling these elements {\bf c}lustered {\bf r}egularly {\bf i}nterspaced {\bf s}hort {\bf p}alindromic {\bf r}epeats, or CRISPR.

\begin{figure}[ht!]
\begin{center}
\includegraphics[width=0.7\textwidth]{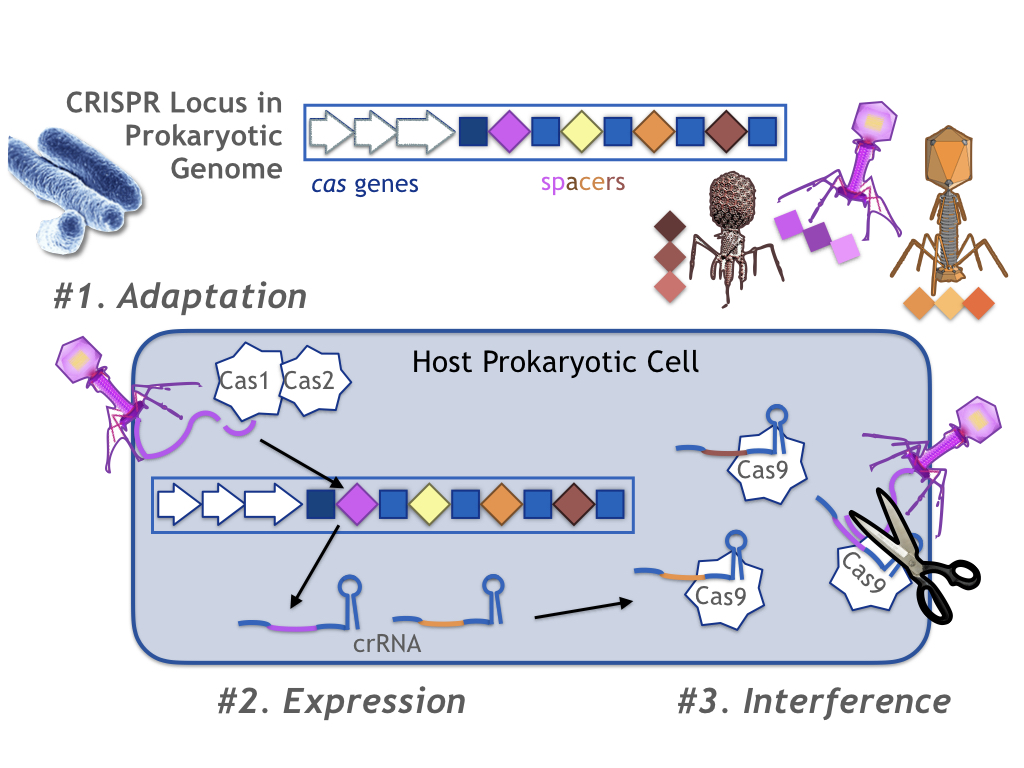}
\caption{Researchers have designated three stages of CRISPR-Cas immunity in a host bacteria or archaea cell, mediated by Cas proteins.  New spacers against viral, plasmidic, and other mobile genetic element foes are acquired during adaptation.  These spacers are transcribed during the expression stage into guide sequences (crRNAs) that team up with a DNA nuclease Cas protein or complex to protect the host from attack by a matching invader during the interference stage.}
\label{fig:3mechanisms}
\end{center}
\end{figure}

Over the following decade, it became clear that CRISPR constituted an adaptive genetic immune system, and experimental studies with \emph{Streptococcus thermophilus} and  \emph{E. coli}  uncovered three distinct phases of adaptation~\cite{27}, expression~\cite{28}, and interference~\cite{32} that are mediated by {\bf C}RISPR-{\bf as}sociated (Cas) proteins.  See Figure~\ref{fig:3mechanisms}).  During adaptation, the CRISPR acquires spacers from protospacer regions in virulent bacteriophage that the prokaryote encounters in its immediate environment and incorporates these into the CRISPR locus immediately adjacent to the leader repeat sequence.  During expression, small {\bf CR}ISPR {\bf RNA}s (crRNA) for each spacer are cleaved from a long, multiunit {\bf pre}cursor crRNA (pre-crRNA) transcription of the locus.  During interference, crRNAs guide the Cas proteins to specifically cleave matching DNA sequences of invading bacteriophage.  Note that we have distinguished between Cas proteins and \emph{cas} genes via capitalization and italics.

Initial comparative-genomic analyses of CRISPR loci and \emph{cas} genes led researchers to interpret the system as a prokaryotic version of the eukaryotic {\bf RNA i}nterference (RNAi) immune mechanism~\cite{39}.  However, a fundamental difference between the two systems is that CRISPR's guide crRNA targets DNA, not mRNA as in eukaryotic RNAi~\cite{37}.  Additionally, these two systems do not share any proteins or noncoding components~\cite{165}, and while long-term immunity can be acquired by eukaryotic RNAi defense systems, it is not heritable~\cite{77}.  The CRISPR spacers, conversely, are inherited by the prokaryotic progeny.

In 2010 as researchers' understanding of the structure and function of these CRISPR-Cas systems was still unfolding, the earliest mathematical models were constructed to study the selection pressure for CRISPR systems~\cite{90} and for the acquired spacers~\cite{83}.  Later models looked further into implications of CRISPR-Cas for the coevolutionary dynamics of host and phage genomes~\cite{7}.  The CRISPR-Cas systems provide a wealth of interesting concepts to study, including coevolutionary dynamics, feedback loops, specificity, efficient organization of the locus and Cas machinery, and horizontal gene transfer.

In this review, we provide an overview of the building blocks of CRISPR-Cas in different species in Section ~\ref{sec:CRISPRClassificationandComponents}.  We discuss the dynamics and diversity of spacers in Section~\ref{sec:MolecularMemoryCassettes}.  We consider the role of and effect on horizontal gene transfer in Section~\ref{sec:HorizontalGeneTransfer}.  We review the mediating characteristics of CRISPR's high specificity in Section~\ref{sec:Specificity}.  We analyze CRISPR evolution and prevalence in Section~\ref{sec:EvolutionAndAbundanceOfCRISPRLoci}.  We enumerate the regulating factors for optimized utilization of CRISPR in Section~\ref{sec:CostAndRegulationOfCRISPRActivity}.  We describe the non-immunological uses of CRISPR-Cas by the host cell in Section~\ref{sec:NonImmunologicalMechanisms}.  We list biotechnology applications in Section~\ref{sec:ApplicationsInBiotechnology}.  We conclude in Section~\ref{sec:Conclusion} with an outlook on how the physics community can contribute to this growing field of study.

\section{Three stages of immunity}
\label{sec:CRISPRClassificationandComponents}

\begin{figure}[ht!]
\begin{center}
\includegraphics[width=0.8\textwidth]{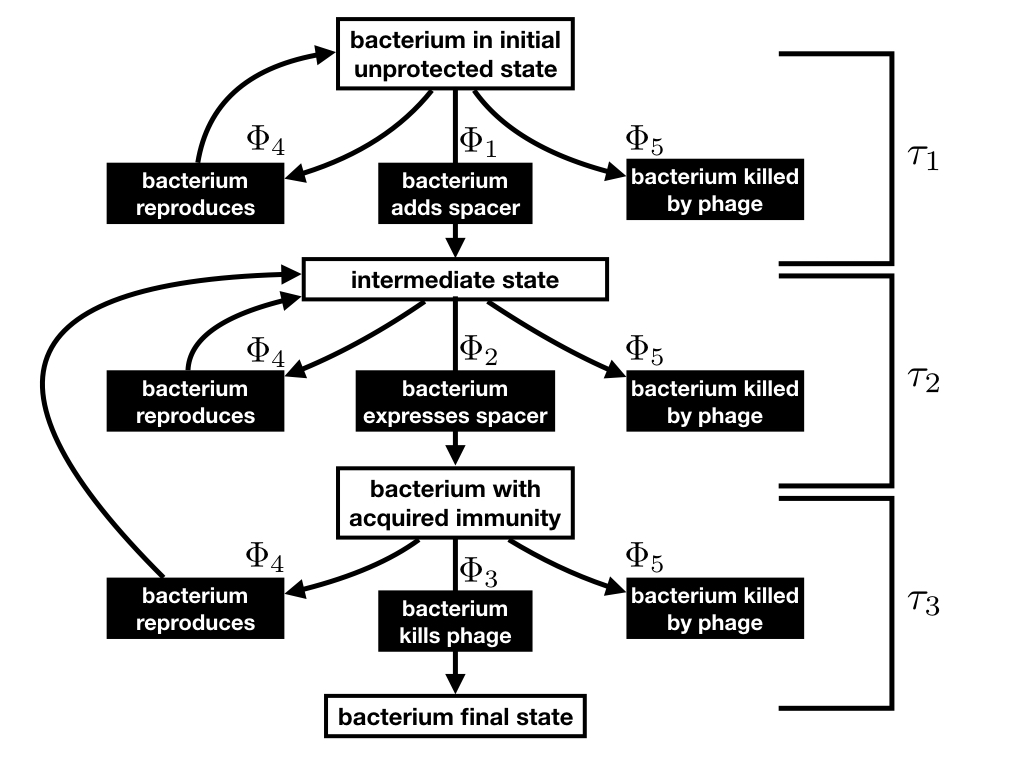}
\caption{A Markov model for CRISPR adaptation, expression, and interference.  The transition event rates $\Phi_{i}$ depend on the characteristic timescales for adaptation by spacer acquisition $\tau_{1}$, expression $\tau_{2}$, and interference $\tau_{3}$.}
\label{fig:markov}
\end{center}
\end{figure}

Our current understanding of the genetic adaptive mechanisms of CRISPR-Cas systems is that they follow a Markov chain.  We describe the transition events for the state change of a combined bacteria and phage system~\cite{8} or for the state change of an individual bacterial cell, as seen in Figure~\ref{fig:markov}.  Each event in the Markov process occurs with a probability proportional to the event's rate $\Phi_{i}$.  In the case where a bacterium begins in an initial state without protection against a particular phage, it must obtain a spacer and express it as a crRNA. If this particular phage strain attacks again, the bacterium uses the crRNA to interfere.  At each state, there is a probability that the bacterium will reproduce or be killed by a phage.   This chain of events could be broken down further to include the probability of bacterium-phage interaction and the probabilities of a lytic or lysogenic phage attack.  The characteristic timescales $\tau_{i}$ of each stage of immunity are still not entirely understood (see Sec~\ref{Timescaleofspacerexpression}).

All CRISPR-Cas systems follow the same pattern of acquiring spacers, transcribing these into mobile surveillance crRNAs, and utilizing the crRNAs as templates to interfere with matching sequences that are attempting to enter the cell.  However, there is a wide variety of components and procedures followed to carry out these mechanisms.  Structural and biochemical studies have provided a detailed genetic and molecular understanding of the unique and conserved components and mechanisms.  The known CRISPR-Cas systems are characterized into two overarching classes based on the type of effector module used during interference and are subsequently divided into six types and 22 subtypes based on signature protein families and distinctive loci architectural features~\cite{111,213}.  Accordingly, Class 1 systems are those that use a multi-subunit crRNA-effector complex, whereas Class 2 systems use a single subunit crRNA-effector protein.  Makarova and colleagues provide a useful ``SnapShot'' of the most up-to-date classification~\cite{215,214}.  The organized classification scheme provides a framework for identifying common threads among the immune systems of different microbial species and calls attention to those systems that are especially distinct.

\subsection{Adaptation}
\label{sec:Adaptation}

Cas1 and Cas2 are the proteins responsible for processing DNA substrates into spacer precursors, and they are highly conserved among different CRISPR types~\cite{100}.  Cas1 is an essential endonuclease during spacer acquisition, and while Cas2 also has DNA/RNA cleavage capability, this is not believed to be important to Cas2's role~\cite{226}.  Cas1 alone can integrate only a small number of spacers, and Cas2 alone can not integrate any.  High performance acquisition therefore requires Cas1 and Cas2 together~\cite{116}.  Non-CRISPR proteins such as RecBCD in  \emph{E. coli}  and Csn2 in \emph{S. thermophilus} may also be recruited for adaptation~\cite{44}.  

\begin{figure}[ht!]
\begin{center}
\includegraphics[width=0.7\textwidth]{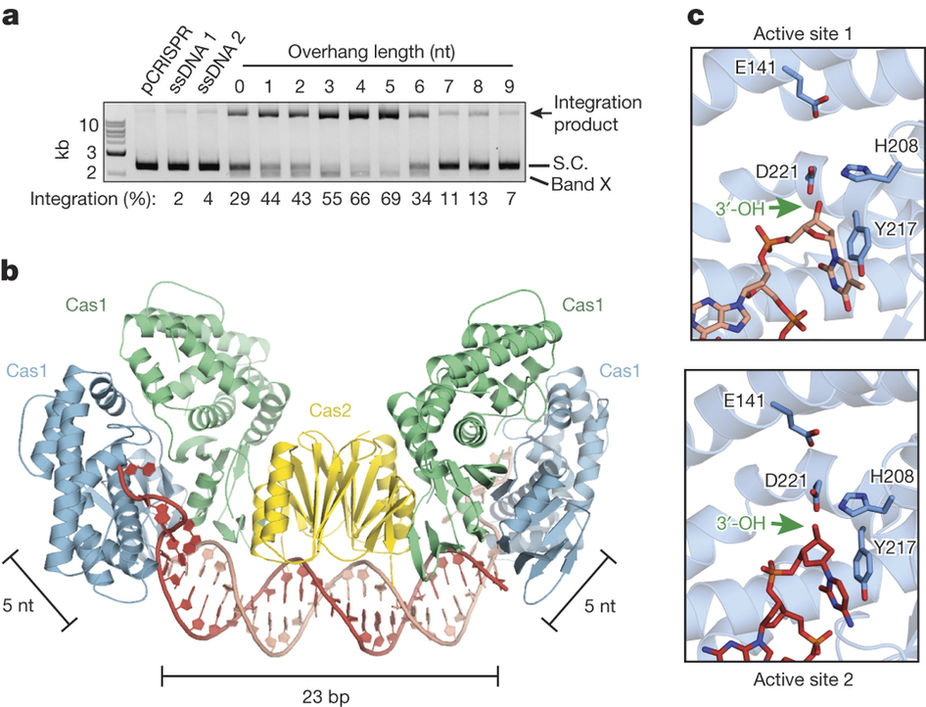}
\caption{The Cas1:Cas2 architecture and active site positioning. (a) \emph{In vitro} integration reaction experiments reveal that the protein complex prefers protospacer DNA with five overhanging nucleotides on each 3' end, as evidenced by the darkest band of integration product.  (b) The protospacer DNA (red) bound to Cas1:Cas2 spans almost the complete 33-nt length of the protein complex.  (c) There are two active sites in the outer Cas1 subunits that facilitate binding to the protospacer's 3' ends.  Reused with permission from~\cite{115}.}
\label{fig:cas1cas2}
\end{center}
\end{figure}

X-ray crystal structures of the  \emph{E. coli}  Cas1:Cas2 complex bound to its protospacer DNA substrate have been used to further uncover the structural basis for foreign DNA capture and integration~\cite{115}.  See Figure~\ref{fig:cas1cas2}.  The protein complex consists of two Cas1 subunits on either end of a Cas2 dimer and two regions in the center, called the `arginine clamp' and the `arginine channel,' used to stabilize the protospacer.  It has a curved binding surface that stretches the length of the spacer to be integrated, acting as a molecular ruler to preserve uniformity of the CRISPR locus sequence architecture.  The ends of the protospacer were splayed to allow its nucleophilic 3'-OH ends to enter channels leading into Cas1 active sites.  The optimal 33-nucleotide substrate for this type of CRISPR system was found to be double stranded DNA with a central 23-bp helical region, flanked by five single-stranded nucleotides on each 3' end.  This requirement for a 33-bp protospacer length was not followed as strictly \emph{in vitro} as it was \emph{in vivo}~\cite{116}.  Non-specific sequence binding resulted from the phosphodiester interactions between the protospacer and Cas proteins~\cite{115}.  The specificity of sequence selection is described in Section~\ref{sec:Specificity}.

The above mentioned studies have mainly focused on `naive' spacer acquisition, in which the CRISPR collects spacers from an invader it has not yet encountered.  If the spacer no longer completely matches the targeted protospacer, either due to spacer degeneration or protospacer mutation, the CRISPR may engage in `primed' acquisition, in which it collects new spacers from an invader it may have been immune to in a previous generation.  For this type of adaptation situation, Cas3 has been shown to be important~\cite{44}.  Priming is discussed in more detail in Section~\ref{sec:IncompleteTargetRecognition}.

The molecular mechanisms of spacer integration, along with the roles of Cas1 bound to Cas2 and the leader-proximal end of the CRISPR array, have been explored \emph{in vivo} by examining induced acquisition of up to three spacers by the Type I-E system in  \emph{E. coli} ~\cite{175}.  Site-specific, staggered nicking occurred at both strands of the leader-proximal repeat, and the 5'-ends of the repeat strands were joined with the 3'-ends of the incoming spacer.  \emph{In vitro} work showed that during integration, Cas1 catalyzed a nucleophilic attack at the 3'-OH ends of the DNA substrate~\cite{116}.  The primary sequence of the first DNA repeat is crucial for having the CRISPR array nicked to incorporate a new spacer~\cite{175}.  Only one repeat sequence is required for spacer integration to occur, and the efficiency of integration is not dependent on whether the array has only this one repeat or a full cassette of repeats and spacers~\cite{100}.  The leader sequence must be at least 60 bp in length~\cite{100}, and it appears to have a cruciform structure joined by AT-rich regions because Cas1:Cas2 preferentially integrates spacers adjacent to this type of sequence hallmark~\cite{116}.  For CRISPR-Cas systems that utilize a {\bf p}rotospacer {\bf a}ssociated {\bf m}otif (PAM), this PAM sequence defined the orientation of the new spacer during integration~\cite{44}, and generally Cas1:Cas2 oriented the 5' G as the first nucleotide~\cite{116}.
	
\subsection{Expression}

After acquisition, spacers are transcribed as crRNAs to guide effector modules for invader interference.  Long {\bf pr}ecursor {\bf crRNA} (pre-crRNA) transcripts are processed from the CRISPR array and cleaved into the individual crRNAs by Cas enzymes in most systems and by an endogenous endoribonuclease in Type II systems~\cite{44}.  Interestingly, a streamlined functional architecture for crRNA maturation was discovered in the \emph{Neisseria meningitidis} Type II-C locus~\cite{158}.  Typically CRISPR-Cas systems contain an external promoter, but here the terminal 9 nucleotides of each CRISPR repeat carried its own promoter element, allowing pre-crRNA transcription to initiate independently in each spacer.  Algorithms have been developed to determine the coding strand that will be transcribed into mature crRNAs and predict crRNA array orientation~\cite{183,91}.  Repeat sequence and mutation information are input, without the need for prior knowledge of type, subtype, class, or superclass of array or repeat, and a variety of factors are considered, such as repeat sequence motifs and biological knowledge of CRISPR evolution.  Understanding the direction of crRNA transcription paves the way for further identification of CRISPR features, including locus conservation, leader regions, target sites on protospacers, and PAMs.

\emph{In vitro} assays and structural analysis of the \emph{Pseudomonas aeruginosa} Type I CRISPR-Cas system was used to understand the protein-RNA interactions that allow Cas6f (formerly Csy4) to recognize and selectively cleave pre-crRNA into crRNA~\cite{137}.  Cas6f processes pre-crRNA with high sequence specificity by recognizing the hairpin element of the CRISPR repeat sequence and cleaving immediately downstream of it.  A 2'-hydroxyl in the nucleotide group immediately upstream of the cleavage site halts cleavage.  The protein has a two-domain architecture to mediate these interactions with RNA, with three important, but not required, residues.  Cas6f is structurally similar to the crRNA biogenesis proteins Cas6e (formally Cse3 or CasE) and Cas6 in \emph{Thermus thermophilus} and \emph{Pyrococcus furiosus}, respectively, which suggests that these all may have come from a single ancestral endoribonuclease enzyme.  For organisms in the \emph{Sulfolobale} order, which typically contain Type I and Type III loci~\cite{177}, a {\bf C}RISPR DNA repeat {\bf b}inding {\bf p}rotein (Cbp1) involved in regulating the production of pre-crRNA transcripts also exists~\cite{202}.  Deleting or over-expressing the \emph{cbp1} gene in \emph{Sulfolobus islandicus} brought about a large reduction or large increase in pre-crRNA yields, respectively.  It is possible that this protein minimizes interference from transcriptional signals that may be carried on A-T rich spacer sequences.  The \emph{cbp1} gene is suggested to have other cellular functions, since it is not physically linked to the CRISPR locus.

Type II systems uniquely express an additional `{\bf tra}ns-activating' {\bf crRNA} (tracrRNA) to anchor the guide crRNA to its single protein effector module Cas9 and position the crRNA for subsequent DNA interference~\cite{44}.  The tracrRNA was discovered from RNA sequencing of \emph{Streptococcus pyogenes} and had a 24-nucleotide complementarity to pre-crRNA repeat regions~\cite{30}.  The tracrRNA binds to Cas9 (formally Csn1) to facilitate base-pairing with the pre-crRNA's repeats and promotes pre-crRNA cleavage into crRNA by an endogenous endoribonuclease III (RNase III).  Though non-Cas, the RNase III is an additional pathway to mature crRNA equivalent to Cas6, Cas6e, and Cas6f.  The other CRISPR Class 2 system arrays, for Types V and VI, are processed into mature crRNAs without a trans-activating crRNA, as tracrRNA is not needed to mediate DNA interference~\cite{12}.

\subsection{Interference}

The crRNA-guided DNA recognition stage of immunity is carried out by either a single Cas protein, \emph{e.g.}, Cas9, or a multicomponent {\bf C}RISPR-{\bf as}sociated {\bf c}omplex for {\bf a}ntiviral {\bf de}fense (Cascade)~\cite{44}.  The specificity of target recognition by these ribonucleoproteins is described in detail in Section~\ref{sec:Specificity}.  The Cascade complex (formally Cmr complex) is composed of a variety of Cas proteins, generally with a static backbone of six Cas7 (formally Cmr4 or CasC) units~\cite{135,79}.  The Type I complex in  \emph{E. coli}  is 405 k-Da with five additional proteins: one Cas8e (formally Cse1 or CasA), two Cas11 (formally Cse2 or CasB), one Cas5 (formally CasD), and one Cas6e (formally Cse3 or CasE)~\cite{79}.  The Class 1 Cascade modules have architectural similarities amongst themselves, and could have evolved from a common ancestor.  However, they are phylogenetically distinct, most evidently in that Type III surveys target DNA in a PAM-independent process~\cite{44}, and Type I must recruit Cas3 for target cleavage~\cite{135}.  

Interestingly, the ``CRISPR Craze'' in genomic engineering~\cite{D}, discussed in Section~\ref{sec:ApplicationsInBiotechnology}, has centered around the use of Cas9 from Class 2, Type II systems, though these are the rarest in nature~\cite{22}.  They are found only in about 5\% of bacteria genomes and rarely in the presence of other CRISPR types.  The attraction arises since these systems have a single multidomain interference protein that performs all of the endonuclease activities required for site-specific DNA targeting.  In nature, Cas9 is guided by the dual tracrRNA:crRNA module, though CRISPR-Cas-based genetic engineering typically makes use of the Cas9 interference machinery and a {\bf s}ingle {\bf g}uide {\bf RNA} (sgRNA), which is a chimeric sequence engineered from a crRNA and a stabilizing tracrRNA~\cite{36}.

Analogous to Cas9's role in Type II systems, Cas12a (formally Cpf1) is a RNA-guided DNA nuclease responsible for target interference in Type V CRISPR-Cas systems~\cite{12}.  Of the 16 Cas12a-family proteins, many exhibit strong structural conservation of the direct repeats.  The \emph{Francisella novicida} Cas12a contains a single RuvC-like endonuclease domain that cleaves target DNA with a 5-nt staggered cut distal to the 5' T-rich PAM.  Distinct from Cas9, Cas12a does not contain an HNH domain nor does it use a G-rich PAM.  Cas12a was shown to be sensitive to mismatches between the crRNA and target DNA in the first eight PAM-proximal nucleotides, especially when there were four consecutive mismatches, but it does not make as extensive contact with its crRNA as does Cas9~\cite{112}.

In Class 2, Type VI systems, RNA is targeted by a variant of Cas13, which contains two higher eukaryotes and prokaryotes nucleotide-binding domains for RNA cleavage~\cite{169}.  The Cas13a1 (formally C2c2) protein in Type VI-A oral bacterium \emph{Leptotrichia shahii} was tested in  \emph{E. coli}, and exhibited successful defense of the cell from an RNA bacteriophage.  Any single mismatches between the crRNA and targeted sequence were tolerated, double mismatches permitted cleavage depending on their location, and triple mismatches did not allow cleavage to occur.  Cas13a1 additionally cleaved non-target RNA after cleaving the targeted strand in what was known as `collateral effect,' causing cell toxicity.  
	
Recently, a computational database mining approach discovered a Class 2 Type VI-B CRISPR locus that uses the interference protein Cas13b to target single stranded RNA, and it expresses two guide crRNAs, a short 66-nt sequence and a longer 118-nt sequence~\cite{216}.  The targeting activity of Cas13b from Type VI-B1 \emph{Bergeyella zoohelcum} and Type VI-B2 \emph{Prevotella buccae} was studied in  \emph{E. coli}.  Targeted sequences typically contained double-sided protospacer flanking sequences, equivalent to the PAM in other CRISPR systems, and in the presence of target RNA, non-target RNA is cleaved due to the collateral effect.  Most fascinating is that this CRISPR-Cas system does not code for Cas1 and Cas2, but it contains two novel Cas proteins Csx27 and Csx28 that regulate Cas13b by respectively repressing and enhancing the effector protein activity.

\section{Molecular memory cassettes}
\label{sec:MolecularMemoryCassettes}

The CRISPR system achieves control over invading phage by incorporating and maintaining a memory of representative pieces of the phage genome. This process by which a bacterium incorporates the protospacer genetic material from a phage as spacers within its CRISPR array is termed adaptation.  Spacer acquisition was first demonstrated in \emph{S. thermophilus} in 2007 \cite{27}, and there are now numerous bacterial CRISPR systems in which spacer acquisition has been observed experimentally \cite{92}.  The adaptive immune response of CRISPR is customized toward a particular foreign invader by utilizing the memory bank of previous encounters \cite{96}.  The evolution of the CRISPR array is generally rapid, on the timescale of days in some cases, and this allows the bacteria to respond to changing pressures of the evolving phage in the environment.

\subsection{Timing and origin of acquired spacers}
\label{sec:Timingandoriginofacquiredspacers}

There are mixed results for the infection conditions that induce adaptation, seemingly dependent on the prokaryotic domain.  In one study of  \emph{Sulfolobus}  archaea, viral DNA replication appeared to be required in order to spark CRISPR spacer acquisition~\cite{189}.  On the other hand, experiments with \emph{S.\ thermophilus} bacteria have shown that encounters with replication-deficient, ``defective'' phage facilitate high spacer acquisition rates~\cite{153}.  This second case is analogous to human vaccination with inactive viruses, which facilitate antibody production for protection from future encounters with an active microbe of the same type.  \emph{In vivo} studies of the \emph{Sulfolobus solfataricus} CRISPR-Cas system found that CRISPR-Cas targeting was independent of the presence of a promoter in front of the protospacer sequence~\cite{204}.  That is, transcription of the targeted gene did not affect immunity~\cite{204,189}.

In all CRISPR systems except Type III, the PAM outside of the protospacer is crucial for spacer acquisition~\cite{63}, as well as target interference.  In general, a number of factors influence selection of new protospacers for incorporation as spacers into the CRISPR array~\cite{92}, however there appears again to be a dependence on the prokaryotic domain.  Spacer recruitment in  \emph{Sulfolobus}  archaea from foreign DNA showed a bias towards plasmid-like sequences versus viral sequences~\cite{177}.  For \emph{S. islandicus} and \emph{S. Solfataricus}, protospacer selection from invading genomes was random and non-directional~\cite{189}.  Additionally, in an early sequence analysis of the spacers in \emph{Crenarchaeal acidothermophile} archaea, the distribution of protospacers along the viral and plasmid genomes and the DNA strand specificity appeared to be uniformly random~\cite{80}.

Conversely, bacteria spacer acquisition appears biased based on genomic location and effectiveness of the derived spacer.  In one study of \emph{S.\ thermophilus}, the most frequently targeted phage sequences were those that were transcribed early during infection~\cite{31}, which would theoretically allow the CRISPR-Cas system to rapidly interfere with the phage during infection and recover before the infection became too severe.  In another \emph{S.\ thermophilus} study, there was a strong and reproducible bias of spacer recruitment from five broad phage genome regions, but to a first approximation, this bias was not related to nucleotide sequence, melting temperature, GC content, single-strand DNA secondary structure, or transcription pattern~\cite{63}.   A metagenomic analysis of \emph{Synechococcus} bacterial CRISPRs from Yellowstone 
hot springs found several spacers matching lysin protein genes and lysozyme enzymes in the phage, which attack the bacterial cell wall late in the phage infection cycle, causing cell lysis and the release of progenies~\cite{85}.  Inactivation of these enzymes halts the spread of the phage and is beneficial for the bacterial population.  These biases in protospacer location and effectiveness may therefore be an evolved bacterial response to the pressure of phage infection.

RNA spacers can also be acquired from RNA phage, as seen in Type III-B CRISPR-Cas systems~\cite{134}.  Spacer integration from single-stranded RNA, single-stranded DNA, and double-stranded DNA was investigated in \emph{Marinomonas mediterranea} by fusing Cas1 to a reverse transcriptase.  In this study, as with several other studies of Type III systems, no sequence signature such as a PAM was associated with protospacer incorporation.  When Cas1 was functional but the reverse transcriptase was not, spacer acquisition occurred only with DNA sequences.  When both proteins were functional, an experiment in which a novel mutation was introduced in the target RNA and observed to propagate to the spacer confirmed that RNA spacers were being acquired.  Thus, a mechanism of reverse-transcribing the integrated RNA spacer to convert it into a DNA equivalent of the crRNA guide was inferred for interference.

\subsection{Experimental studies of spacer diversity}

The diversity of spacers has been a long-standing quantity of experimental interest.  Bacterial population resistance was shown to greatly increase as within-population spacer diversity increased due to the~\cite{196}.  It has been observed that more active loci typically have more diverse spacers overall~\cite{33}, even among loci within a single strain~\cite{176}.  Diversity along the CRISPR array is highly linked to the fact that spacer insertion typically occurs at the leader-proximal end of the CRISPR array~\cite{170}.  Some sporadic spacer insertions have been observed at inner locus positions in \emph{E.\ coli}, albeit to a much lesser extent than those occurring at leader repeat position~\cite{176,175}.  In one of the six CRISPR loci of \emph{Sulfolobus solfataricus}, spacers were preferentially incorporated at repeat number four~\cite{170}.  It is unclear if these instances are due to erroneous recruitment or binding of Cas1-Cas2 integrase complexes to internal repeat sequences.  Despite these exceptions, polarized growth of the CRISPR locus in the leader-proximal end means that spacer order yields an exact chronological record of virus encounter.

The degree to which spacers in a population match coexisting targets is generally locus position dependent~\cite{60}.  There is selective pressure for retention of useful, older spacers that match more of the dominant phage genotypes~\cite{201,171,189}.  For instance, in a metagenomic study of archaeal, bacteria, and viral populations in Lake Tyrrell from 2007 to 2010, the spacers and their targeted viruses were stable over days, and these spacers were generally retained for one to three years~\cite{65}.  Experiments have furthermore observed more variability at the leader-proximal end of the locus, while leader-distal end spacers are highly conserved in bacteria~\cite{33,171,60,64}.  The leader-distal spacers of the locus appear to experience a loss of diversity in order to provide an evolutionary advantage against persistent viruses.  An exhaustive analysis of all currently known spacers in archaeal genomes showed that here too spacers targeting common viruses were located further away from the leader sequence~\cite{93}.  Interestingly, a five-year metagenomic study of \emph{Leptospirillum} group II bacteria in biofilm found that the conserved leader-distal spacers did not perfectly match the dominant phage DNA~\cite{60}.  It is possible that these degenerate spacers are useful to the host for primed adaptation (see Section~\ref{sec:IncompleteTargetRecognition}).

Individual CRISPR loci have active gain and loss of spacers, suggesting that each strain is exposed to different phage during its life history~\cite{176,171,60,129}.  The \emph{Leptospirillum} group II bacteria have spacers common to the species group located at the leader-distal site, population-specific spacers towards the middle, and unique, single-copy spacers at the leader-proximal site~\cite{56}.  Spacers located in equivalent positions among a species or population, as well as specific leader-distal clonality, have additionally been observed in other sequences analyses~\cite{56,176} and in long term metagenomic studies~\cite{64}.  In cases where most of the spacers are shared between two species, such as with \emph{Mycobacterium bovis} and \emph{Mycobacterium tuberculosis}, it was suggested that these species encountered many of the same phage~\cite{129}.

\subsection{Modeling spacer diversity in the CRISPR locus}

\begin{figure}[ht!]
\begin{center}
\includegraphics[width=0.5\textwidth]{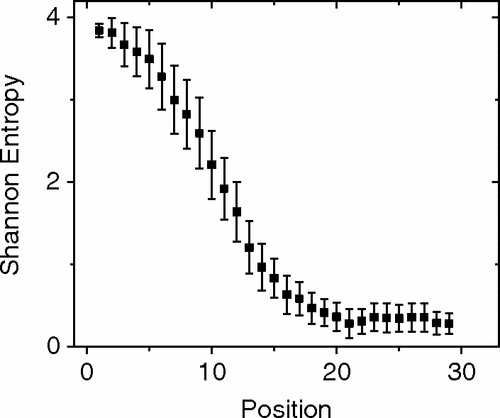}
\caption{The Shannon entropy, calculated with Eq.~\ref{eq:shannon}, measures spacer diversity as a function of position in the CRISPR-Cas locus.  A population dynamics model of bacteria and phage reveals spacer diversity decreases from the leader-proximal ($x$-axis origin) to the leader-distal end.  This agrees with experimental observations.  Reprinted with permission from~\cite{83}.}
\label{fig:diversity}
\end{center}
\end{figure}

Though the CRISPR array of spacers provides a record of the phage challenges that the bacteria have faced, this record is convolved with the effects of selection on the utility of the retained spacers.  In one of the first theoretical studies of CRISPR, a population dynamics model was used to explain the experimental observation that the leader-proximal end of the spacer array is more diverse than the leader-distal end~\cite{83}.  In this model, old spacers were dropped when CRISPR reached a certain length, 30 spacers in this case, to avoid infinite growth, and the heritable locus was copied to two daughter cells after bacterial division.  The system of mean field equations for the densities of bacteria $x$ with spacers $i$ and $j$ and phage $v$ with protospacer $k$ that interact with each other at a rate $\beta$ were
\begin{align}
\frac{\textrm{d}x_{i,j}}{\textrm{d}t} &= a x_{i,j} - \beta\sum_{k \ne i,j}v_{k}x_{i,j} + \beta\gamma\sum_{m}x_{j,m}v_i \\
\frac{\textrm{d}v_{k}}{\textrm{d}t} &= r v_k - \beta\sum_{i,j}x_{i,j}v_k(\delta_{i,k}+\delta_{j,k}).
\label{eq:meanfield}
\end{align}
The first term in the bacteria population density equation is exponential growth at a rate $a$ in the absence of phage, the second term is the loss of bacteria due to lack of protection from a spacer matching the infecting phage, and the third term represents spacer gain events that occur with a probability $\gamma$.  In the phage density equation, the first term is exponential phage population growth at a rate $r$ in the absence of CRISPR bacteria, and the second term represents bacteria-phage encounters that degrade the phage if the bacteria has a matching spacer.  The diversity $D$ for spacer position $i$ was calculated by the Shannon entropy as
\begin{equation}
D_i=-\sum_{k}p_i(k)\textrm{ln}p_i(k),
\label{eq:shannon}
\end{equation}
where $p_i(k)$ is the probability to have sequence $k$ at position $i$.  Figure~\ref{fig:diversity} shows that, even with extension of the model to include the possibility of virus mutation, the diversity of spacers was found to decrease with position from the proximal end.  Those spacers that matched the largest fraction of the phage population were selected for initially, and this reduced-diversity set of spacers was shifted to the leader-distal end as additional, more diverse spacers were incorporated to the leader-proximal end. The resistant bacteria gained a selective advantage, meaning those spacers providing resistance remained fixed in the population.  

\begin{figure}[ht!]
\begin{center}
\includegraphics[width=0.75\textwidth]{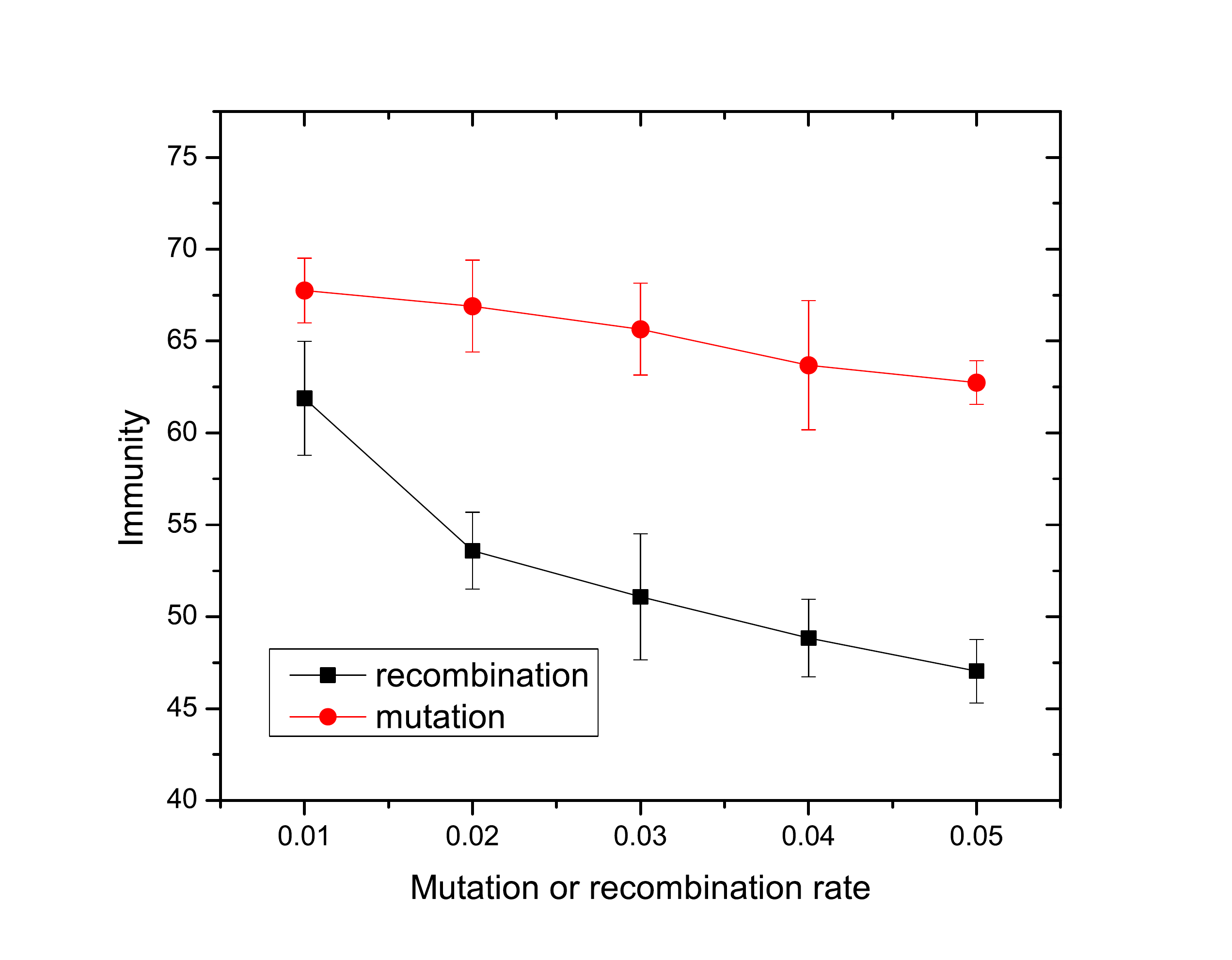}
\caption{The effect of phage mutation and recombination on CRISPR recognition.  Immunity quantifies the ability of a CRISPR spacer to effectively recognize a phage threat and stop infection.  The CRISPR has a mismatch tolerance of two basepairs, therefore it can catch escape mutants that have one or two point mutations.  Recombination is the more successful mechanism of phage evasion of CRISPR recognition.  Reprinted with permission from~\cite{8}.}
\label{fig:recombination}
\end{center}
\end{figure}

A refined model included the effects of protospacer recombination in the phage as well as other types of spacer deletion mechanisms in the bacteria~\cite{8}.  Phage were able to avoid recognition by CRISPR using point mutation, which led to mismatch between crRNA sequence and that of invading phage, and recombination, which incorporated mutations that increased fitness and increased the chance of a mismatch that would allow the phage to escape CRISPR recognition.  Recombination that integrated multiple point mutations was shown to be a more effective evolution mechanism of phage escape than point mutation alone (see Figure~\ref{fig:recombination}).  Spacer diversity again decreased towards the leader-distal end due to selection pressure.

When there are functional differences in effectiveness of different spacers, the observed distribution of spacers in the CRISPR array is a convolution of the effects of selection and ease of acquisition.  A population dynamics model of CRISPR was used to explore how spacer effectiveness and ease of spacer acquisition allow bacterial and phage populations to co-exist, oscillate, or be driven to extinction~\cite{50}. In the absence of functional differences, protospacers were acquired and inserted into the CRISPR proportional to their acquisition probability. With a single protospacer in phage and a constant rate of spacer loss in bacteria, oscillations in bacterial and virus populations were seen due to successful infections of the wild-type bacteria, which led to an increase of phage, followed by an exponential increase of protected bacteria that have effective spacers and decrease of phage, and then the creation of more susceptible bacteria due to space loss.  With multiple protospacers that differed in their ease of acquisition, bacteria developed a diverse locus of spacers, and with multiple protospacers that differed in their effectiveness, a less diverse, specialized spacer distribution appeared.  Often the steady state bacterial population did not reach the maximum capacity due to presence of virus, but if it occurred, the phage were usually driven to extinction.

Spacer diversity was also shown to be important to the survival of bacterial strain populations in a numerical model of spatially distributed bacteria and phage~\cite{87}.  A range of spacer numbers was investigated to reflect the natural diversity of CRISPR array lengths, with strains containing between 0 to 20 spacers.  It was found that spacer diversity diminished for older spacers.  Additionally, the spacer usage frequency fell off rapidly after a short distance along the CRISPR array.  The average number of spacers evolved to be between 20 to 30.

\subsection{Effects of spacer acquisition and deletion rates}

\begin{figure}[ht!]
\begin{center}
\includegraphics[width=0.55\textwidth]{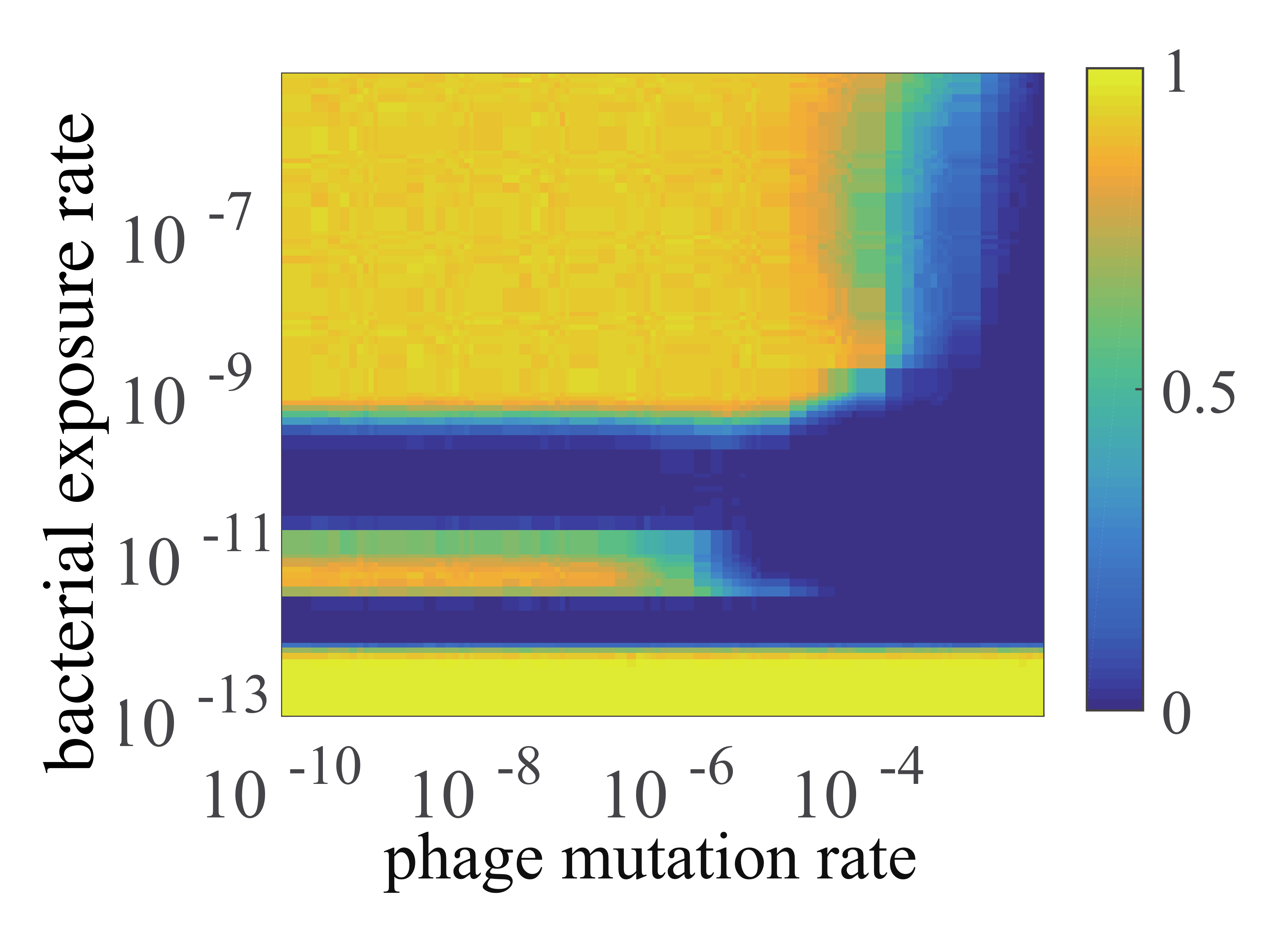}
\caption{Extinction probability of phage in coevolutionary model with CRISPR bacteria in which there are five phases and four transitions.  The probability of bacteria acquiring spacers $\gamma$ is 5 x $10^{-4}$ and the rate of deleting spacers $\gamma{}'$ is 10$^{-5}$ min$^{-1}$.  The phage burst size $\rho$ is 100, the lysis rate $r$ is 0.025 min$^{-1}$, the phage decay rate $d$ is 0.001 min$^{-1}$, the bacteria growth rate $a$ is 0.005 min$^{-1}$, the bacteria carrying capacity $q$ is 10$^{7}$ mL$^{-1}$, the initial bacteria density $x_0$ is 5 x 10$^{6}$ mL$^{-1}$, and the initial phage density $v_0$ is 5 x 10$^{7}$ mL$^{-1}$.  The bacteria have a maximum locus length $L$ of 6 spacers, the cost $\alpha$ of having CRISPR immunity is 0.1 per spacer, and the number of available protospacers $N_p$ is 30.  The total volume of the system $V$ is 10$^{-3}$ mL.  Bacterial exposure rate $\beta$ is in units of mL $\cdot$ min$^{-1}$ and phage mutation rate $\mu$ is in units of min$^{-1}$.  Reprinted with permission from~\cite{201}.}
\label{fig:extinction}
\end{center}
\end{figure}

There have been a few theoretical models that have explored how the rate of spacer acquisition affects locus diversity and the coevolving bacteria and phage populations.  For instance, in a strain-level model of the coevolution of bacteria and phage, strain diversification was tied to the spacer acquisition rate~\cite{86}.  In a stochastic model, small rates of phage mutation and varying spacer incorporation and deletion rates led to a nonclassical bacteria and phage coevolution phase diagram (Figure~\ref{fig:extinction})~\cite{201}.  In particular, at low rates of spacer deletion $\gamma{}'$, \emph{e.g.}, $10^{-5}$ min$^{-1}$, the phage population size depended in a nonmonotonic way upon the spacer acquisition and deletion rates.  The ability of phage to mutate or recombine their protospacers results in a reentrant phage-bacteria phase diagram that is distinct from the classical predator-prey phase diagram.  When the phage mutation rate $\mu$ is low, there are five phases and four transitions in the phage extinction probability.  At very low bacteria exposure rates $\beta$, phage extinction probability is high due to the infected bacteria lysis rate $r$ and phage burst size $\rho$ not being greater than the phage decay rate $d$.  

The mean field equations for this stochastic model with
 resistant bacteria $z$,
susceptible bacteria $x$,
infected bacteria $y$, and phage $v$ are 
\begin{eqnarray}
\frac{d z}{d t} &=& - \gamma' z + c^z z + \beta \gamma v x
\label{eq:transition1a}
\\
\frac{d x}{d t} &=& \gamma' z + c^x x - \beta v x
\label{eq:transition1b}
\\
\frac{d y}{d t} &=& \beta (1-\gamma) v x - r y
\label{eq:transition1c}
\\
\frac{d v}{d t} &=& \rho r y - \beta v (x + z) - d v
\label{eq:transition1d}
\label{eq:transition1}
\end{eqnarray}
Here the replication rate is given by $c^x = a [1-(x+y+z)/q]$ 
 for carrying capacity $q$, and $c^z = c^x / (1+\alpha)$ for
spacer cost $\alpha$.
At low mutation rates, for the parameter regime of Figure \ref{fig:extinction},
we find by setting Eq.\ \ref{eq:transition1c} to zero
and examing the growth rate of Eq.\ \ref{eq:transition1d} that
there is a transition at
$\beta^*_1 = d/[q (\rho{}-1)]
\approx 10^{-12}$ mL $\cdot$ min$^{-1}$,
after which phage have a very high survival probability due to their replication rate exceeding their decay rate.  If the bacteria did not have a CRISPR immune system, there would be no further phase transitions, and phage would survive for all exposure rates 
$\beta > 10^{-12}$ mL $\cdot$ min$^{-1}$.  In either case, the phage population increases with increasing exposure rate.  With CRISPR bacteria, an increasing phage population triggers an increase in the number of bacteria with spacers.  
At steady state, the combined bacterial populations nearly
reach the carrying capacity $q$ if they are not extinct,
and $\gamma' \gg c^x$.
If the phage exist, the steady state of
Eq.\ \ref{eq:transition1b}
implies concentrations
of $v^* = \gamma' z^* / (\beta x^*)$
and
the steady state of Eq.\ \ref{eq:transition1c}
implies
$y^* = \gamma' z^* / r$
since $\gamma \ll 1$.
This leads eventually to a second transition, because
Eq.\ \ref{eq:transition1d} expresses the
condition $(\rho-1) \beta  x > d $ for a positive
growth rate of the phage.
The value of $x$ increases quite rapidly due to bacterial growth, nearly to $q$, but
then decays as a result of spacer acquisition due to
phage pressue.  The value to which $x$ decreases changes non-linearly
with $\beta$ so that for $\beta^*_2 > \beta^*_1$ there
is an extinction of phage at this intermediate time.
 If instead the rate of spacer loss is higher in the bacteria, \emph{e.g.}, $\gamma{}' = 10^{-4}$ min$^{-1}$, the susceptible
bacteria are not driven to as low a value at intermediate times, and
this phage extinction phase would not exist.  Even for a low rate of spacer loss
the phage extinction phase eventually disappears as the
encounter rate is increased.
After the intial phage burst drives all bacteria to acquire spacers,
susceptible bacteria are created by spacer loss, with
 $x \approx \gamma' q t$.  In this regime $\beta q \ll d$, and
so the mean field equations become
\begin{eqnarray}
\frac{d y}{d t} &=& \beta v \gamma' q t - r y
\label{eq:transition3a}
\\
\frac{d v}{d t} &=& \rho r y - d v
\label{eq:transition3b}
\label{eq:transition3}
\end{eqnarray}
The phage extinction is a result of stochastic effects, and
we locate the transition approxmately as when there is a single
virus particle in the system of volume $V$, 
 $v V \approx 1 $.  This occurs at a minimum of $v(t)$ at time $t^*$, 
so $\rho r y = d v$,
thus $y(t^*) = d / (\rho r V)$.    We further approximate
$v(t) \approx v(0) \exp(-d t)$ 
to find $t^* \approx (1/d) \ln [V v(0)]$ and
in Eq.\ \ref{eq:transition3a}.
We solve Eq.\ \ref{eq:transition3a} 
to find when $d \ll 1/r$ and $t^* \gg 1/r$
that $y(t) \sim \beta \gamma' t q v(0) \exp(-d t)/ r$.
Setting $y(t^*) = d/(\rho r V)$ 
we find
 $\beta^*_3 = d^2/\{\rho \gamma' q \ln [V v(0)] \}$, which is about $10^{-11}$
 mL $\cdot$ min$^{-1}$.
 During this phase, phage grow back and survive with high probability.  If instead the bacteria could acquire an infinite number of spacers while phage still have a finite number of protospacers, this phase would not exist.  
The fourth transition occurs when the exposure rate is so
high that all bacteria are driven to  absorb spacers.
The susceptible bacteria each produce $\rho$ phages.  This transition
occurs at a minimum of $x(t)$, so $x = \gamma' z / (\beta v - c^x)$.
In this regime $\beta v \gg c^x$, and when we set $x V = 1$
to examine the vanishing of susceptible bacteria, we
find $\beta^*_4 =\gamma' q V / [\rho x(0)]$, or
about $2 \times 10^{-10}$ mL $\cdot$ min$^{-1}$.
 Phage have a low survival probability at high exposure rates because bacteria rapidly recognize threats and extinguish the phage population before they have lost all of their spacers.  
The regions of phage survival and extinction in this nonclassical phase diagram are more sensitive to the bacterial rate of losing spacers, as increasing the probability of acquiring spacers only increases phage extinction at high phage mutation rates.

A third model showed how spacer diversity depended on the overall bacterial acquisition rate when there were encounters with a single phage that had multiple possible protospacers~\cite{50}.  Large acquisition probabilities led to a broader spacer diversity distribution, whereas smaller acquisition probabilities led to selection for and greater abundance of the most effective spacers.  Interestingly, a rapid increase in the spacer uptake rate increased the likelihood of spacers that self-target the bacterial host genome.  This theoretical result agreed with an experiment in which an engineered Cas9 in \emph{S. pyogenes} led to increased spacer acquisition but also increased autoimmunity~\cite{198}.  That is, even assuming a constant CRISPR array length, an increased rate of acquisition meant a single bacterium would incorporate a greater number of spacers, and so there was a greater cumulative probability that one of those spacers would activate an autoimmune response.  Autoimmunity has been observed in species containing wild type CRISPR machinery as well~\cite{33}.  Another interesting result from this experiment is that Cas9, or at least this mutated version of the protein, appears to play a previously unrecognized role in spacer acquisition~\cite{198}.

\subsection{Timescale of spacer expression}
\label{Timescaleofspacerexpression}

Whether the CRISPR is able to incorporate protospacers from an active phage infection in time to protect that bacterium against the infection is unclear.  That is, do all three mechanisms of adaptation, expression, and interference occur fast enough to protect an individual from a newly encountered invader? It has been suggested that the completion of these three mechanisms may not actually happen on a fast enough timescale to interfere with phage replication in a naive phage-infected cell before the cell becomes lethally damaged from the infection~\cite{153}.  It was proposed that the source of protospacers is from phage that are defective due to mutations, DNA damage, faulty genome packaging, or degradation by another host defense mechanism.  This exposure to defective phage is a form of vaccination and imbues the cell with future protection against infection by non-defective phage.

To understand the timescale of the expression phase, during which CRISPR-Cas transcript processing takes place, a minimal model was developed \cite{118}.  With half lives of pre-crRNA and crRNA that are on the order of minutes and hours, respectively, the model showed that a fast decay of pre-crRNA leads to increased production of crRNA.  A very strong increase in processing rate of the enzyme that catalyzes pre-cRNA to crRNA processing led to fast, non-specific loss of pre-crRNA. Due to Le Chatelier's principle, this reduced concentration of intermediate substrate significantly enhanced crRNA generation.  These results echoed those of an experiment in which an increase of pre-crRNA to crRNA was achieved by significant over-expression of the Cas enzyme that catalyzes this transcription~\cite{174}.

\section{Horizontal gene transfer}
\label{sec:HorizontalGeneTransfer}

Horizontal gene transfer (HGT) is the exchange of genetic material between individuals not necessarily of the same species.  Pangenomic analyses, which consider core versus non-core genes among different strains, have shown that HGT plays a role in the stability and flexibility of conserved and functionally essential genomic structures of prokaryotic genomes~\cite{125}.  It has been shown theoretically that HGT, coupled to modularity, accelerates the rate of evolution in a population of individuals on a rugged fitness landscape~\cite{231}.  For short times $t$ and a finite number of individuals $N$, the average fitness $F$ in the population increases as
\begin{align}
\langle{}F(t)\rangle{} &= 2L + \lambda_{1}t+\lambda_{2}t^2, \label{eq:HGT} \\
\lambda_{1} &= 2L\left(1-\frac{1}{N}\right), \nonumber \\
\lambda_{2} &= -\frac{4L^2}{N}\left(1-\frac{1}{N}\right)-4\mu{}L\left(1-\frac{1}{N}\right)-2\nu{}L\left[\left(1-\frac{1}{K}\right)(1-M)\left(1-\frac{4}{N}\right)+\frac{1}{N}\right]\left(1-\frac{1}{N}\right), \nonumber
\end{align}
where each individual has a genetic sequence composed of $L$ sites in $K$ modules with a modularity $M$, and the sites had a mutation rate $\mu$ and modules have a HGT rate $\nu$~\cite{230}.  Note that modularity couples to the horizontal gene transfer rate, as $(1-M)$ appears together with $\nu$.  Modularity increases the fitness at short times.  The increase in fitness due to modularity is proportional to the rate of HGT.

The CRISPR-Cas system is physically modular on several scales, from the level of individual spacers up to the entire system being considered a module.  As we will show, there is evidence in support of HGT of whole CRISPR-Cas systems across prokaryotes.  However, the integrity of CRISPR hinders further HGT from occurring within the locus or with other parts of the genome.  This complex relationship with HGT has led to both the evolution and the stability of CRISPR systems of different species.

\subsection{Acquisition of CRISPR loci and spacers}

CRISPR families were identified through analysis of sequences and system architecture, including CRISPR repeats, spacers, leader sequences, and \emph{cas} gene content~\cite{78,8}.  These families did not necessarily correlate with the classical phylogenetic tree~\cite{66}.  This is evidence of the CRISPR-Cas system being propagated by inter-genus and inter-species HGT events, followed by further evolution.  A large-scale phylogenetic analysis of \emph{cas} genes suggested CRISPR loci are propagated between cells on megaplasmids~\cite{172}.  A ``total evidence'' tree based upon phylogenetic analysis of the complete CRISPR locus, revealed that CRISPRs and \emph{cas} genes are a form of mobile genetic element that disseminates via HGT as a single module.  About 15\% of the CRISPR-cas loci were on megaplasmids rather than on the host chromosome, and many of these loci were also present in distantly related genomes.  These results indicate that the CRISPR-cas locus has been passed by means other than vertical transmission, such as HGT or conjugation.

Genomic data from \emph{E.\ coli}, \emph{P.\ aeruginosa}, \emph{S.\ agalactiae}, and \emph{S.\ thermophilus} strains were analyzed with an inference algorithm to determine which CRISPR spacers in bacterial strains were received from recombination events~\cite{53}.  Without recombination, it is expected that order will be conserved at the leader-distal end and diversified at the leader-proximal end.  The analysis looked for order divergence events, \emph{i.e.}, additional patterns of spacer content similarity between strains that would have been introduced from lateral spacer transfer.  These events are observed as shared segments followed by different segments towards the leader-distal end.  This has similarly been seen in \emph{Leptospirillum} group II bacteria CRISPR arrays, where there are abrupt transitions in the loci for population-specific spacer regions~\cite{56}.  It was estimated that only about 10\% of \emph{S. thermophilus} strains received spacers from recombination events, and similar results were found for other examined strains~\cite{53}.  These results demonstrate recombination, but also suggest that recombination in these species of bacteria is likely not especially advantageous for rapidly improving phage resistance.  A  bioinformatic analysis of CRISPR loci in \emph{Mycobacterium} revealed similar repeats and \emph{cas}1 genes among genera that are orders apart, such as  \emph{M.\ tuberculosis} and \emph{Bifidobacterium adolescentis}, and suggested horizontal gene transfer of the CRISPR locus would explain these findings~\cite{129}.

\emph{Sulfolobus}  archaea have very extensive and diverse viral, plasmidic, and other mobile genetic element foes, which explains their highly extensive and diverse CRIPSR loci of Type I and III~\cite{177}.  Even so, there is evidence in support of whole CRISPR-Cas module transfer between organisms within \emph{Sulfolobus} CRISPR-Cas systems~\cite{123}.  Additionally, an analysis of archaeal species spacers showed the presence of archaeal chromosomal genes in CRISPR loci, including those that must have been acquired from inter-genus and inter-species gene transfer events~\cite{93}.  One mechanism for inter-genus or inter-species transfer is acquisition of a chromosomal region by a natural plasmid via recombination, which is transferred by conjugation to a new cell.  The CRISPR-Cas system would then recognize this plasmid as foreign and copy some of the genetic material as spacers.  Another mechanism is one in which a virus defectively packages a portion of its host's DNA during infection into a ``transducing particle,'' which could enter an archaeal cell and trigger CRISPR-Cas adaptation.  It is thought, however, that more significant barriers exist to transfer of the system between archaea and bacteria~\cite{123}.  A possible mechanism is one in which spacers matching eukaryotic and bacterial genes could have been acquired from non-specific archaeal natural competence and subsequent CRISPR-Cas activation~\cite{93}

\subsection{CRISPR-Cas restriction of HGT}

\begin{figure}[ht!]
\begin{center}
\includegraphics[width=0.7\textwidth]{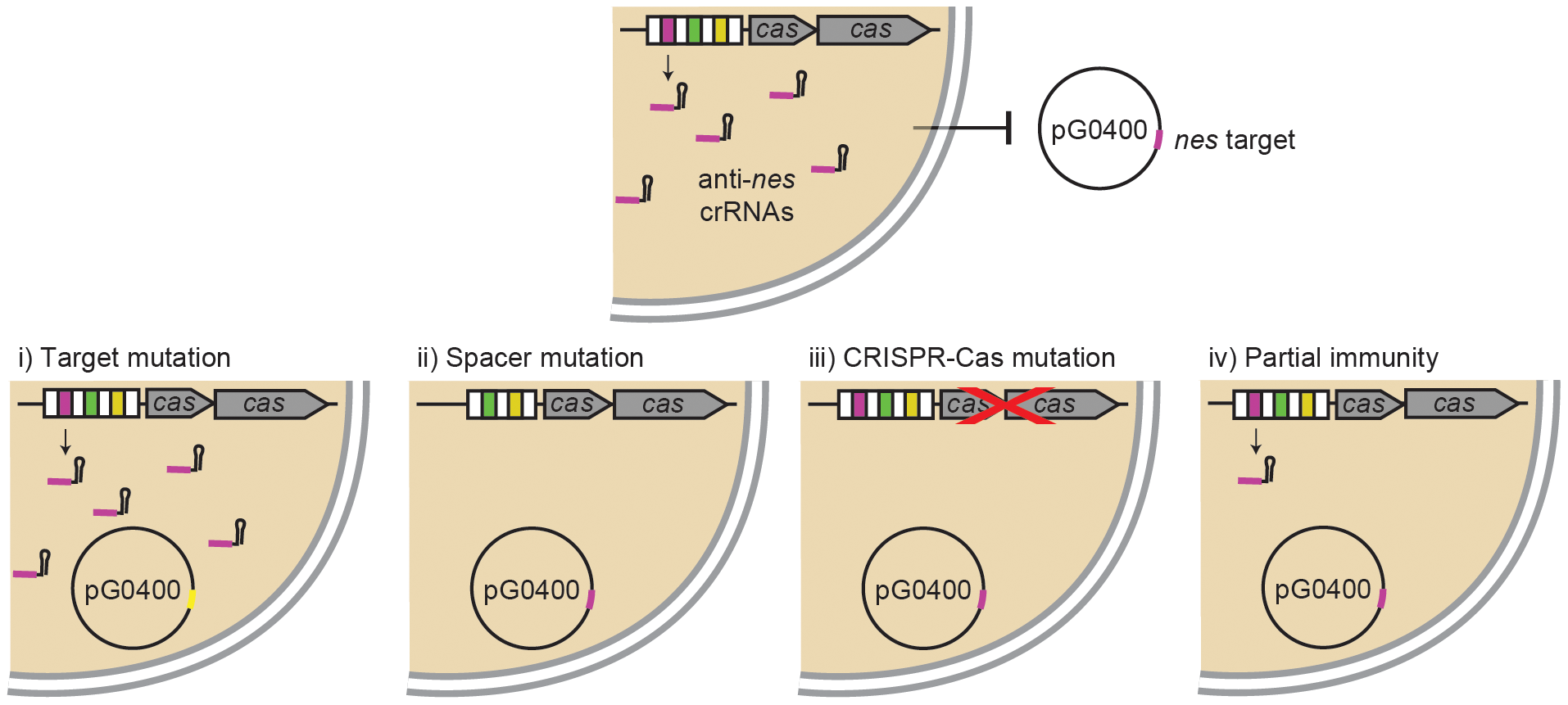}
\caption{If bacteria containing CRISPR are challenged by a plasmid containing a beneficial gene, CRISPR activity is suppressed to allow uptake of the plasmid.  Suppression can occur due to mutation in the target protospacer, mutation in the host's spacer, mutation in the \emph{cas} genes that render CRISPR ineffective, or a weakened CRISPR-Cas response.  Reprinted with permission from~\cite{102}.}
\label{fig:HGT}
\end{center}
\end{figure}

Many multi-drug resistant and virulent isolates have gained their resistance genes from genetic elements acquired during viral invasion, called prophage, or from plasmids.  In cases where this HGT is essential or highly beneficial to an organism, CRISPR-Cas constitutes a fitness cost, and suppression of CRISPR activity is crucial to the survival of these organisms~\cite{102}.  See Figure~\ref{fig:HGT}.  Computer modeling and experiments indicate loss of CRISPR-Cas loci in the presence of an environment containing prophage or plasmids that increase the host's fitness~\cite{51}.  Indeed, most naturally occurring human bacterial pathogens that survived antibiotic selection lack CRISPR-Cas loci~\cite{152,102}.  This selection pressure leads to evolution such that CRISPR-Cas systems are in continuous flux.  They can be lost when they block lateral transfer of beneficial genes and gained when there is phage infection pressure.

Genomic sequencing has provided insight into how CRISPR limits the virulence of clinical strains versus deadly, food-borne strains~\cite{199} and limits the presence of drug resistance genes~\cite{154,152,23}.  The active CRISPR loci in \emph{Cronobacter sakazakii} clinical strains, capable of causing disease, had significantly fewer spacers than those in food-borne strains~\cite{199}. These fewer spacers in clinical strains explain why they had more prophage than food-borne strains and were more virulent.  Rapid gain and loss of prophage and CRISPR spacers caused dynamic evolution of \emph{C. sakazakii}.  Similarly, genomic analysis revealed a high inverse correlation between \emph{Enterococcus faecalis} species containing CRISPR-cas and those with antibiotic resistance genes, suggesting antibiotic use unintentionally selects for strains that compromise genome defense \cite{154,152}.  Likewise, the CRISPR system in \emph{Staphylococcus epidermidis} inhibits this bacteria's ability to develop antibiotic resistance, whereas \emph{Staphylococcus aureus} has increased virulence due the scarcity of CRISPR loci~\cite{23}.

Multiple bacterial experimental studies have shown how CRISPR prevents HGT through the direct targeting of DNA in \emph{Staphylococci}~\cite{37}, \emph{Streptococcus pneumoniae}~\cite{105}, \emph{Neisseria}~\cite{158}, and  \emph{E. coli} ~\cite{162}.  For instance, the transfer of a particular plasmid conferring antibiotic resistance occurs easily from \emph{S. aureus} to \emph{S. epidermidis} in the absence of CRISPR~\cite{37}.  When \emph{S. epidermidis} was engineered to contain a CRISPR locus with a spacer targeting this plasmid, plasmid transfer only occurred if the targeting spacer was deleted.  In another experiment, \emph{S. pneumoniae} CRISPR loci were engineered to contain a spacer for the capsule gene, a pneumococcal virulence factor~\cite{105}.  In the presence of the engineered CRISPR, HGT was mostly blocked and \emph{in vivo} infection in mice was unsuccessful.  Furthermore, as CRISPR caused cell death in cells infected with the capsulated strain, this supported the possibility of engineering mobile CRISPR systems to target antibiotic resistance or virulence in infectious bacteria for patient care.  Additional studies have confirmed that CRISPR-Cas systems affect emergence and virulence of human bacterial pathogens through HGT barriers and gene expression modulation~\cite{152}.

\subsection{Persistent HGT}

Some researchers question how likely CRISPR-Cas systems are to collect spacers against beneficial plasmids in nature~\cite{136}.  There are indeed exceptions to the negative correlation between CRISPR-positive bacteria and pathogenicity discussed in the previous section.  For instance, the virulent and multi-drug resistant \emph{Clostridium difficile} contains multiple CRISPR repeat regions, with several actually located in the prophage~\cite{168}.  An interesting phage mechanism that could account for these exceptions is the use of anti-CRISPR proteins to provide a loophole for HGT to occur.  Phage that attack \emph{P. aeruginosa} encode five distinct families of CRISPR-inhibiting proteins that block Type I-F and four families that block Type I-E CRISPR systems~\cite{192}.  These phage, therefore, carry their own shield against CRISPR-Cas interference.  The anti-CRISPR proteins bind various parts of the Cas complex and regulate lateral gene transfer by allowing foreign DNA to bypass recognition by CRISPR-Cas.  In a similar manner, a \emph{P. aeruginosa} pathogenicity island found in a highly virulent clinical isolate contains an anti-CRISPR homologue~\cite{223}.  This anti-CRISPR homologue is likely what allows transfer of the pathogenicity island between \emph{P. aeruginosa} by conjugation~\cite{192}.

\section{Specificity}
\label{sec:Specificity}

The specificity of the CRISPR-Cas machinery is a high concern for comprehensive immunity in prokaryotes and for avoidance of off-target activity in biotechnology applications.  At the basic level, CRISPR must distinguish between itself and foreign DNA so that it does not integrate self-DNA as a spacer nor mistake the spacers in its CRISPR locus as threats.  Some amount of cross-reactivity is beneficial because requiring exact matches between the crRNA and target DNA would disadvantage prokaryotes that are facing phage that may mutate their protospacers in an attempt to avoid CRISPR recognition~\cite{221}.  The balance between having a weak response to self antigens and a strong response to non-self antigens is a universal issue in immune system dynamics.  The human immune system, for example, has evolved with selection for antibodies that are not cross reactive on average to avoid autoimmune disease~\cite{186}.

Interestingly, the CRISPR-Cas systems evolved to have modular and hierarchical specificity.  The three modules of target recognition are the protospacer associated motif (PAM), the first 8-12 protospacer nucleotides adjacent to the PAM known as the seed region~\cite{20}, and the remainder of the roughly 30-bp protospacer.  The tolerance threshold for the number of mismatches that leads to target interference or no target interference has been studied both experimentally and in theoretical models.  Mismatches in each of the modules hold different weights for the Cas proteins' ability to recognize target DNA.  Additionally, there are certain instances when an intermediate recognition of target DNA uniquely regulates the CRISPR-Cas response to bind without cleavage, and this phenomenon is discussed in more detail in Section~\ref{sec:IncompleteTargetRecognition}.

\subsection{Cas specificity and conformational changes}

The PAM is an important aspect of invader recognition by Cas proteins during spacer acquisition and target interference.  This 2-5 bp motif is generally not contained in the protospacer nucleotides and varies among different CRISPR systems and organisms~\cite{227}.  Structural and biochemical studies of the Cas proteins used in adaptation and interference have helped to shed light on their PAM specificity.  During adaptation, the Cas1 dimer in the Cas1:Cas2 complex functions as a sequence-specific pocket that recognizes the PAM-complementary sequence~\cite{226}.  As mentioned in Section~\ref{sec:Adaptation}, the precise length of the sequence that is cleaved for spacer integration is determined by the length of the Cas1:Cas2 complex.  For interference, it is believed that the Cas protein and guide crRNA complex scans putative invader DNA for a PAM, and upon finding one, the complex initiates binding of its crRNA to the sequence downstream of the PAM~\cite{191}.  

\begin{figure}[ht!]
\begin{center}
\includegraphics[width=0.8\textwidth]{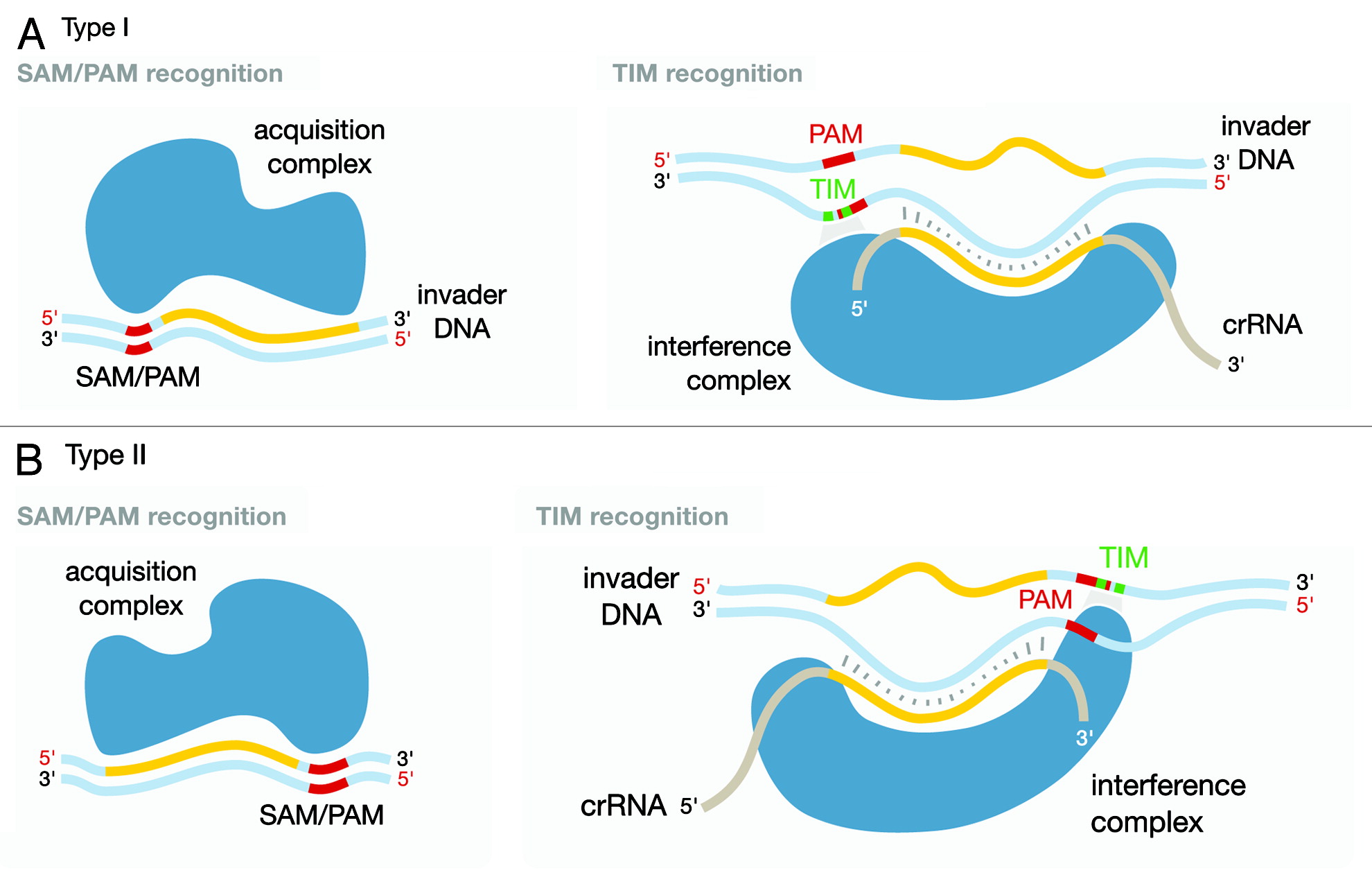}
\caption{There are putative separate protospacer adjacent motifs (PAMs) recognized by the adaptation and interference Cas proteins.  The suggested acronyms are spacer acquisition motif (SAM) and target interference motif (TIM).  In both types, the SAM is recognized by the Cas1:Cas2 complex.  (A) In Type I CRISPR-Cas systems, the TIM is recognized on the crRNA-complementary DNA strand by the Cascade complex.  (B) Conversely, the TIM in Type II systems is recognized by Cas9 on the strand non-complementary to the crRNA.
Reprinted with permission from~\cite{227}.}
\label{fig:PAMs}
\end{center}
\end{figure}

Cryo-electron microscopy of the Type I Cascade and Types II Cas9 revealed that, though their structures were fundamentally different with no apparent evolutionary connection, they were mechanistically similar as they both had specific domains for PAM recognition and facilitated the seed interaction between complementary regions of crRNA and target DNA~\cite{20}.  For instance, Cas9 has a carboxyl-terminal domain that was identified to be responsible for the PAM interaction.~\cite{21}. The PAM recognition loops in the Cas9 of different organisms were structurally divergent, probably to account for the distinct guide crRNA and PAM specificities.~\cite{20}.  Since the molecular mechanisms of adaptation and interference differ, it has been suggested that a separate sequence motif is recognized in these two steps~\cite{227}.  Figure~\ref{fig:PAMs} \emph{i.e.} shows the {\bf s}pacer {\bf a}cquisition {\bf m}otif (SAM) and the {\bf t}arget {\bf i}nterference {\bf m}otif (TIM).  More TIMs are recognized than SAMs, at least in Type I systems~\cite{232}, which is a possible mechanism for limiting the probability of acquiring self-targeting spacers.

\emph{E. coli}'s Type I-E Cascade composition and structure have been studied through single particle electron microscopy to understand the physical mechanism of CRISPR surveillance of invader DNA and subsequent binding~\cite{79}.  Cascade had a sequence-specific manner of recognizing doubled-stranded DNA targets that relied on R-loop formation in which, after crRNA base-paired with the complementary DNA strand, the non-complementary DNA strand was displaced, forming an R-shaped loop.  The crRNA:targetDNA complex was tightly bound in 5-nt segments, since crRNA has 6-nt interval kinks that cannot basepair~\cite{20}.  Upon binding to the DNA target, Cascade changed conformation from resembling a seahorse with a curled up `tail' to having less prominent `nose' and `neck' features~\cite{79}.  Cascade-mediated cleavage of the target DNA did not occur, confirming this CRISPR type requires Cas3 for cleavage.  Cas3 recruitment is dependent on specific binding between the crRNA and target DNA~\cite{222} and therefore on Cascade's subsequent conformation change~\cite{79}.  Interestingly, Cascade binds to non-target DNA in a mechanism entirely controlled by the presence of Cas8e in the Cascade complex~\cite{79}.  This non-specific interaction between Cascade and DNA presumably makes target scanning more efficient and enhances sequence-specific DNA localization.

The Type III Cascade complex of \emph{Thermus thermophilus}, which targets single-stranded RNA, was studied through cryo-electron microscopy to understand the target-bound and unbound states~\cite{135}.  The central, double-helical core of the unbound complex was composed of a Cas7 backbone, whose geometry remained unchanged in the bound state.  Rod-shaped segments protruded for engagement with the single-stranded RNA target.  In the bound state, the Cas subunits were rearranged to expose the crRNA and form an elongated channel to accommodate the crRNA:target duplex.  The bound RNA target was then distorted by thumb-like domains for cleavage.  The Type III CRISPR-Cas systems in the archaeal genus \emph{Sulfolobus} are characterized by the additional presence of Cas10, possibly involved in nucleic acid targeting~\cite{177}.

The crystal structure of the Type II \emph{S. pyogenes} Cas9 has been extensively studied alone, in complex with sgRNA, and bound to target DNA in order to shed light on its structure, conformational changes, target surveillance, and PAM recognition~\cite{21,159,72}. The Cas9 bound to a 98-nt sgRNA and 23-nt target DNA exhibited a bilobed architecture, termed a target recognition lobe and nuclease lobe~\cite{21}.  The negatively charged sgRNA:targetDNA was accommodated in a positively charged groove at the interface of the two lobes.  The recognition lobe, which was specific to Cas9 and appeared to be conserved across the Cas9 families, was responsible for binding the sgRNA and target DNA.  The nuclease lobe contained HNH and RuvC nuclease domains positioned for cleavage of the complementary and non-complementary strands, respectively.  The HNH domain was mobile, as it approached the complementary target DNA strand to cleave it through a conformation change in the segment connecting the HNH and RuvC domains.  Alone, Cas9 has an auto-inhibited conformation~\cite{20}, though binding to sgRNA triggers a conformational rearrangement of Cas9 to prepare it for specific DNA binding~\cite{159}.  The X-ray crystallography of \emph{S. pyogenes} Cas9 was compared to that of \emph{Actinomyces naeslundii}, and they showed similar inactive and rearranged conformations sparked by the presence of sgRNA.  Negative stain electron microscopy of Cas9:sgRNA:DNA revealed the rearrangement of the bilobed structure into a central channel.
	
Intramolecular F\"{o}rster resonance energy transfer (FRET) experiments were used to discern the relative orientations of Cas9's catalytic domains during on- and off-target DNA binding of custom targets~\cite{114}.  When the sgRNA was lacking certain features, such as perfectly matching basepairs in the PAM or seed regions, binding to its target DNA substrate did not spark a Cas9 conformational change, and the FRET state was indistinguishable from Cas9:sgRNA alone.  Additionally, an increasing number of mismatches between the sgRNA and substrate DNA led to a diminished HNH conformation change.  DNA cleavage efficiency by CRISPR-Cas9 was shown to depend specifically on the activated conformation of the HNH nuclease domain; for example, substrates with four or greater basepair mismatches led to less of an HNH conformation change and were cleaved slowly, if at all.  Subsequent, coordinated triggering of the RuvC domain nuclease activity was also tied to HNH's conformation change, not HNH nuclease activity, through an allosteric communication pathway.

Most recently, single-molecule FRET identified an intermediate Cas9 conformational state that served as a ``checkpoint'' before the HNH domain transitioned into a catalytically active docked state for target cleavage~\cite{229}.  If the number of mismatches between the guide RNA and the target DNA exceeded a threshold, Cas9 remained in its intermediate conformation.  Additionally, when the guide RNA was truncated, the binding affinity with DNA was lowered, which lowered occupancy of the docked state but also increased the cleavage specificity.

\subsection{Identifying CRISPR-Cas PAMs}

\begin{figure}[ht!]
\begin{center}
\includegraphics[width=0.6\textwidth]{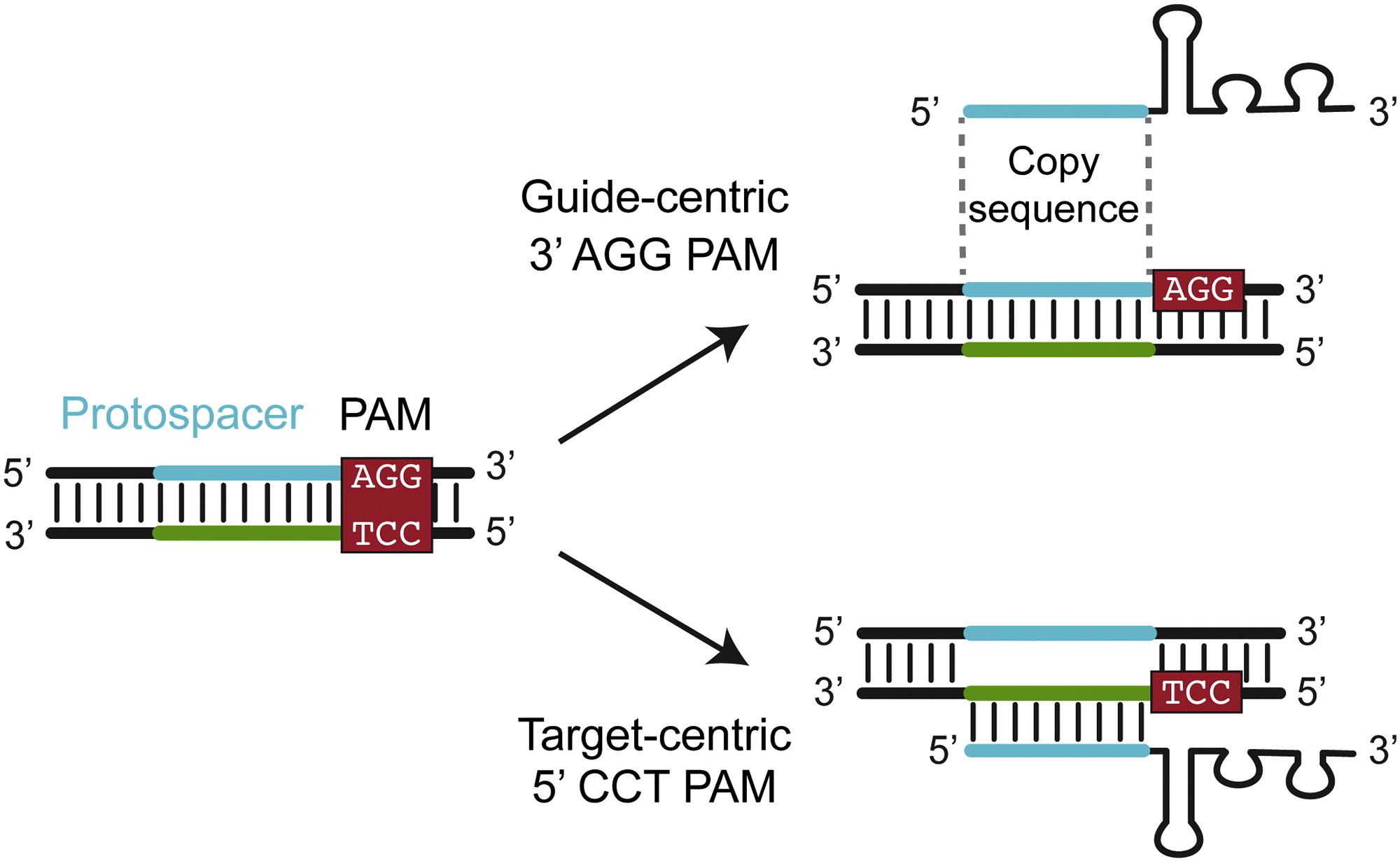}
\caption{A guide-centric reporting scheme for CRISPR-Cas PAMs considers the motif on the target strand that is non-complementary, \emph{i.e.,} identical, to the sgRNA.  On the other hand, a target-centric scheme identifies the PAM from the perspective of the DNA strand complementary to the sgRNA.
Reprinted with permission from~\cite{191}.}
\label{fig:PAMlabel}
\end{center}
\end{figure}

Data-driven analyses have made further progress in identifying the variety of PAMs that different CRISPR types recognize.  To date there has not been a consistent orientation used to report the PAM sequence, with some research groups reporting the PAM and its location relative to the strand that matches the crRNA and others reporting relative to the complementary strand that basepairs with the crRNA, often depending on the type of CRISPR-Cas system~\cite{227}, as was shown in Figure~\ref{fig:PAMs}.  It has been suggested to use a guide-centric orientation scheme, in which the PAM is reported from the non-complementary strand~\cite{191}.  Figure~\ref{fig:PAMlabel} shows a consistent notation that will aid guide RNA design.

A metagenomic study of microbial DNA extracted from an acid mine drainage environment showed that there were consistently more spacer matches to phage with PAMs than to those without~\cite{60}.  During adaptation, experiments have shown that the process of selecting a plasmid sequence to make a spacer is non-random~\cite{138}.  The Type I-E CRISPR loci of \emph{E. coli} was studied when the bacteria were challenged by a plasmid.  Anti-plasmid spacers from protospacers that had an AAG PAM sequence located directly upstream were integrated into the bacteria's loci, leading to protection from the plasmid.  Similarly with interference, experiments have shown that the transformation rate of plasmids into CRISPR-Cas archaea is significantly lower when the plasmids contain PAM trinucleotides~\cite{127}.  The Type I-B CRISPR loci of \emph{Haloferax volcanni} in particular recognize six PAM sequences upstream of protospacers, ACT, TTC, TAA, TAT, TAG, and CAC~\cite{127,232}.  Transformation was restored when the CRISPR-Cas system of the host archaea cell was altered to be defective, most commonly via \emph{cas} gene cassette deletion or mutation, followed by mutation in chromosomal spacer, plasmid protospacer, or PAM.

The importance of the PAM region for successful recognition by CRISPR has lead to recent efforts to improve DNA recognition capabilities in biotechnology applications \cite{191}.  There are a number of methods for determining the set of functional PAM sequences for a particular CRISPR system.  One approach is a BLAST search of metagenomic databases, but a limitation of this method is the availability of sequence information.  Another is to screen for depleted plasmids or sgRNA clearance of phage and the dependence on the presence of a PAM.  Cas proteins have also been engineered to improve CRISPR-Cas system's DNA recognition capability, such as Cas9 recognizing alternate PAMs and Cas1:Cas2 having a relaxed PAM specificity.

\subsection{Self and non-self discrimination}

Without regulation, the CRISPR-Cas system could inadvertently target host genomic material for acquisition, leading to subsequent interference and cell death.  As discussed in the previous sections, spacer acquisition by Type I and II CRISPR systems rely on the presence of a limited number of acquisition PAMs.  The host genetic material will only very rarely match the interference PAM plus spacer. In one study of the importance of the PAM sequence for Type II-A CRISPR-Cas system, it was found that 30 spacers targeted genes of the host genome, but the interference PAM discerned self from non-self recognition~\cite{171}.

Another issue is the crRNA inadvertently matching its spacer in the CRISPR array as though it were part of an invading DNA sequence.  In this case, self and non-self are distinguished by the presence of the repeat sequence adjoining the spacers in the CRISPR array~\cite{94}.  A study with \emph{S. epidermidis} confirmed that extended pairing of the interference machinery and the repeat sequences upstream of the spacers avoids self-targeting.  This mechanism is possible because when the spacer is expressed as crRNA, a few bases of the repeat sequence are also included.  Mismatches between the target sequence and crRNA at specific positions outside of the spacer cue the CRISPR system that the target is foreign DNA.  Conversely CRISPR interference is abrogated when there is complementarity between the crRNA and the nucleotides at positions 2, 3 and 4 upstream of the alleged target.  All CRISPR-Cas loci exhibit the distinctive complementarity of their DNA repeats outside of the spacer sequence to prevent this type of autoimmunity.

As discussed in Section~\ref{sec:Timingandoriginofacquiredspacers}, acquisition of spacers in some CRISPR systems has been linked to phage replication activity.  Experimental work has determined that this activity offers another self and non-self distinguishing mechanism~\cite{97}.  The mechanism revolves around RecBCD, an endogenous bacterial enzyme complex that processes double-stranded DNA break repair.  Firstly, RecBCD readily binds to the end of linear DNA, and since an invading phage in the process of replicating will have open replication forks, RecBCD bind these sequences.  Secondly, during normal host cell repair, RecBCD unwinds the two DNA strands until it reaches the nearest recombination hotspot, called a Chi site.  Recombinatory repair is then carried out by RecA.  If RecBCD binds DNA from a replicating phage, it will degrade the genetic material without stopping due to the phage's lack of Chi sites.  Cas1:Cas2 then takes advantage of this degraded phage DNA substrate for spacer processing and integration.  Therefore, the high number of replication forks on foreign DNA encourages spacer acquisition from foreign DNA, and the high density of Chi sites on self chromosome limits spacer acquisition from self DNA.  Additionally, a lower expression of Cas1:Cas2 leads to a higher specificity for exogenous DNA.

\subsection{Cross-reactivity}

To avoid CRISPR defense, viruses have evolved mechanisms for generating genomic deletions, insertions, and rearrangements~\cite{221}.  Mismatches between the target and the spacer affect the ability of CRISPR to recognize target genetic material, leading to decreased levels of resistance~\cite{204,177}.  Relaxed specificity allows a single crRNA to target a virus that had evolved an escape mutation or to target several related viruses.  Matching between the crRNA and the protospacer in the PAM and seed regions is usually crucial for initial recognition of foreign DNA, because the crRNA uses this seed region to efficiently scan invader DNA for an initial match~\cite{95}.  Conversely, CRISPR-Cas systems are able to recognize viral targets with up to 5 mutations outside of the seed region.   There is also some dependence on the prokaryotic domain, as archaean CRISPR systems generally have a lower specificity than bacterial systems~\cite{178}, with the exception of a strict intact PAM requirement~\cite{123}.

An early theoretical study looked at the minimum number of mismatches needed for the phage to escape via point mutation or recombination~\cite{8}.  When a single mismatch was sufficient for the phage to escape CRISPR recognition, there was little difference between the results from point mutation versus recombination.  However, when two mismatches were required, recombination gave the phage more of a chance to survive, and CRISPR immunity to the recombining phage was lower.  A second model showed that an evolved result of increased cross-reactivity is a reduced diversity required in the optimal immune repertoire of CRISPR spacers~\cite{130}.  While tolerance of mismatches reduces the diversity of spacers needed for protection, the threat of autoimmunity increases.  Indeed, another mathematical model showed the extent of PAM specificity reflected a tradeoff between the host's requirement of a non-negligible probability to acquire diverse spacers to protect itself and avoidance of a high probability of autoimmunity~\cite{54}.

Interestingly, researchers explored the use of a smaller sgRNA for genomic editing that exhibited lowered binding affinity to the target, but also lowered cross-reactivity~\cite{185}.  Profiling of sgRNA off-target activity is discussed in more detail in the following section.  Since naturally occurring CRISPR systems are known to tolerate some alterations in the target sequence, heightened affinity and cross-reactivity from natural-sized sgRNAs are undesirable for biotechnology applications. The decreased length of the sgRNA:targetDNA interface decreased the binding free energy, making the gRNA more sensitive to mismatches.  Indeed, this result echos observations made earlier of the adaptive antibody immune system~\cite{186}.  A computer simulation showed how evolution of antibodies through gene segment swapping and point mutation led to a balance between binding affinity and specificity to avoid autoimmune effects.  A more aggressive immune response resulting from a more thorough search of antibody sequence space leads to more strongly binding antibodies, but also to antibodies with greater cross reactivity.
	
\subsection{Profiling Cas9 off-target activity}

Specificity in biotechnology applications has been of particular concern to ensure that only the target sequence is modified.  There have been several systematic investigations of the binding activity of either a large pool of sgRNAs~\cite{66} or a large array of potential off-target sequences~\cite{182,117}.  The goal is to create data-driven computational models that are predictive of targeting activity and generalized across genes for the design of optimal sgRNAs~\cite{66}.  Whole-genome analysis protocols have been developed, including {\bf g}enome-wide, {\bf u}nbiased {\bf i}dentification of {\bf d}ouble-stranded breaks {\bf e}nabled by {\bf seq}uencing (GUIDE-seq)~\cite{251} and Cas9 nuclease-{\bf di}gested {\bf genome} {\bf seq}uencing (Digenome-seq)~\cite{252}.  These methods are especially important for taking human genetic variation into account when designing specific sgRNAs.

The amount and location of tolerable mismatches that lead to off-target activity have been mapped out.  Two or more mutations occurring in the PAM or the seed region were not tolerated, and multiple mismatches proximal to the seed region reduced sgRNA association~\cite{182}.  Single-base mismatches were more tolerated in PAM-distal, \emph{i.e.}, in the 5' half of the sgRNA, than PAM-proximal regions~\cite{140,66}.  Two base mismatches considerably reduced cleavage activity, and three or more interspaced and five consecutive mismatches usually halted cleavage~\cite{140}.  There were also gene-specific patterns of more effective target sites and sequence features that were found to be more favorable, such as having guanine immediately adjacent to the PAM~\cite{66}.

Several means of optimizing on-target specificity have been identified.  One way to achieve higher specificity is to pair two highly active sgRNAs with Cas9 nickases that each generate a single-stranded DNA break~\cite{67}.  Others include extending the PAM sequence for use with \emph{S. pyogenes} Cas9~\cite{66} and making the Cas9 protein human codon-optimized~\cite{38}.  While extension of the tracrRNA tail of the chimeric sgRNA exhibited an increase in editing efficiency~\cite{140}, a tradeoff was observed between activity and specificity, both in vitro and in cells~\cite{117}. Namely, a shorter, less-active sgRNA was more specific than a longer, more-active sgRNA.  The lower binding affinity from the shorter complementary strands leads to higher specificity and less off-target activity~\cite{185,5}.  Lower concentrations of Cas9:sgRNA also lowers activity, thereby increasing cleavage specificity~\cite{140,5}.  Conversely, high concentrations of Cas9:sgRNA could cleave off-target sites containing mutations near or within the PAM, which usually were not cleaved with lower concentrations~\cite{117}.  Since most single mismatches still achieve high levels of sgRNA:DNA association, genomic editing at locations distinct by at least two bases from the rest of the genome will generally be most precise~\cite{182,67,140}.

Some research has also considered the kinetics of Cas9:sgRNA interactions with target and mismatched DNA strands to obtain a biophysical understanding of the efficiency and specificity of binding and to quantitatively predict off-target activity.  The interactions of catalytically dead Cas9 with a library of potential DNA binding strands was experimentally analyzed to understand the off-target binding potential~\cite{182}.   The effect that one, two, or more mismatches had on association rates was examined in real time with a massively parallel array of mutant targets.  This study demonstrated that mismatches between the sgRNA and DNA at distinct domains of PAM-distal bases modulated different biophysical parameters of association and dissociation.  These results suggested the possibility of using kinetic and thermodynamic tuning of the Cas9:sgRNA interaction with DNA to achieve rapid and specific binding.  

In another study, a quantitative model that encompasses the multi-step process responsible for CRISPR-Cas9-based genome editing and gene regulation was developed~\cite{173}.  The five modeled steps were Cas9 and crRNA expression, Cas9:sgRNA complex formation, diffusion and DNA site selection, reversible R-loop formation with formation of Cas9:sgRNA:DNA complex, and DNA site cleavage.  Several parameters were considered, including sgRNA sequences, DNA superhelical densities, Cas9 and sgRNA expression levels, organism and growth conditions, and experimental conditions.  The study looked at how several factors control outcomes, among them dynamics of Cas9 binding and cleavage at all DNA sites, considering both canonical and non-canonical PAMs.  DNA supercoiling was determined to be a novel mechanism that controls Cas9 binding.  In particular, R-loop formation, from Cas9:sgRNA binding to DNA, negatively supercoils the site's DNA, which positively supercoils adjacent DNA sites, deterring other Cas9:sgRNA from binding there.  The model to predict the sequence-dependent rate $R_{\textrm{binding}}$ for a particular Cas9:sgRNA complex $i$ to bind to DNA sequence $j$ uses
\begin{equation}
R_{\textrm{binding}[i,j]} = p_{[i,j]}R_{\textrm{random walk},i},
\label{eq:bindingRate}
\end{equation}
where $R_{\textrm{random walk}}$ is the contact rate due to molecular diffusion for Cas9:sgRNA complex $i$ and $p$ is the binding probability, 
\begin{equation}
p_{[i,j]} = \frac{\frac{N_{\textrm{target},j}}{N}\exp{(-\Delta{}G_{\textrm{target}[i,j]}/k_{\textrm{B}}T)}}{\sum\limits_{m}\frac{N_{\textrm{target},m}}{N}\exp{(-\Delta{}G_{\textrm{target}[i,m]}/k_{\textrm{B}}T)}},
\label{eq:bindingProbability}
\end{equation}
to one of the total available DNA sites $N_{\textrm{target}}$ with sequence $j$ in the genome of length $N$.  The probability follows a Boltzmann distribution, where $k_{\textrm{B}}$ is the Boltzmann constant and $T$ is temperature. The binding free energy $\Delta{}G_{\textrm{target}}$ is
\begin{equation}
\Delta{}G_{\textrm{target}[i,j]} = \Delta{}G_{\textrm{PAM},j}+\Delta\Delta{}G_{\textrm{exchange}[i,j]}+\Delta\Delta{}G_{\textrm{supercoiling},j},
\label{eq:freeEnergy}
\end{equation}
where $\Delta{}G_{\textrm{PAM}}$ is the free energy originating from the PAM and Cas9 interactions, $\Delta\Delta{}G_{\textrm{exchange}}$ represents the free energy difference between the DNA target bound to its complementary DNA sequence and the sgRNA bound to the DNA target during R-loop formation, and $\Delta\Delta{}G_{\textrm{supercoiling}}$ designates DNA site supercoiling free energy.  The rate $R_{\text{C}}$ for a particular Cas9:sgRNA complex $i$ to cleave a DNA sequence $j$ is then
\begin{equation}
R_{\text{C}[i,j]} = \frac{k_c}{k_c+k_{d_,j}}R_{\text{binding}[i,j]},
\end{equation}
where $k_d$ is the kinetic constant of dissociation of the Cas9:crRNA:DNA complex and $k_c$ is the kinetic constant of cleaving bound DNA.  Off-target binding frequencies were determined across lambda phage and human genomes.  Guidelines were proposed for designing effective genome editing or regulation experiments that minimize off-target activity and maximize on-target binding.  Undoubtedly in some cases, kinetics rather than thermodynamics will dominate off-target activity.  The study of kinetics remains an open problem.

\section{Evolution and abundance of CRISPR loci}
\label{sec:EvolutionAndAbundanceOfCRISPRLoci}

\begin{figure}[ht!]
\begin{center}
\includegraphics[width=0.7\textwidth]{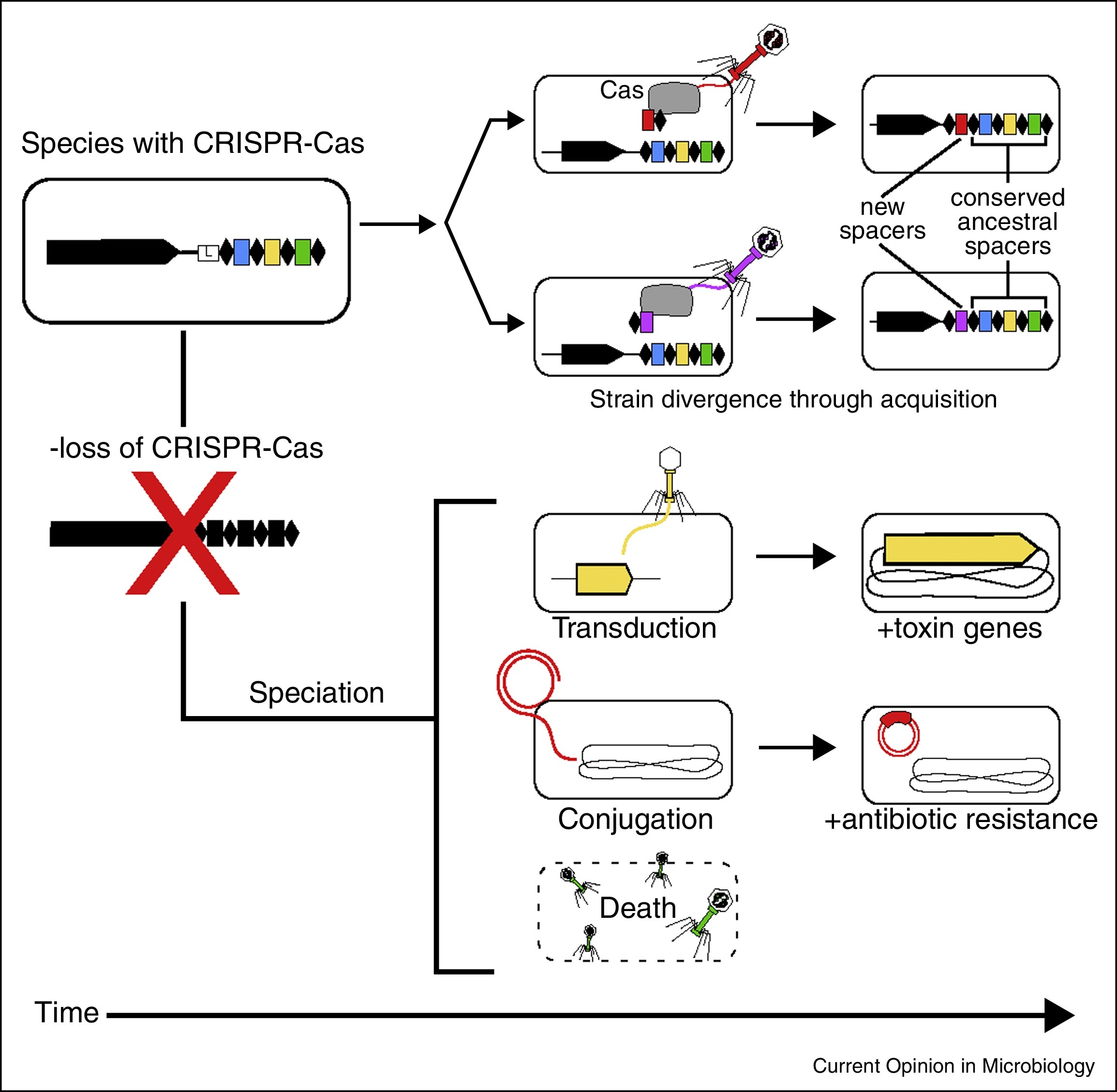}
\caption{Rapid divergence of strains within a species occurs as individual organisms with CRISPR-Cas encounter diverse threats and acquire new, unique spacers.  If CRISPR-Cas is lost, new species can be created through HGT events, such as transduction or conjugation. One example is the loss of CRISPR-Cas in \emph{Streptococcus zooepidemicus} that allowed the bacteria to acquire virulence factors from the uptake of phage genes.  This led to the speciation of a highly pathogenic strain, \emph{Streptococcus equi}.  Reused with permission from~\cite{4}.}
\label{fig:divergence}
\end{center}
\end{figure}

The CRISPR-Cas loci in prokaryotes serve a functional role as protection from phage and plasmid infection.  The evolution of this defense mechanism is therefore based on the fitness advantage that it confers to the host.  General modeling of the evolution of host defense mechanisms has shown that ecological feedback informs evolutionary dynamics, since the ecological time scale is much faster than the evolutionary time scale~\cite{181}.  In the case of CRISPR, ecological feedback to the host from the surrounding phage population and selection pressure for cell survival informs CRISPR-Cas locus evolutionary dynamics.  The divergence of CRISPR-Cas loci in an otherwise homologous prokaryotic population is a result of challenge from invading phage~\cite{4}.  See Figure~\ref{fig:divergence}.  We will also discuss how abundance of CRISPR loci in some individuals of a species and loss of CRISPR loci in other individuals can lead to speciation after evolution of these two groups.

\subsection{Support for a Lamarckian-type evolution}

The dynamic CRISPR-Cas immune system drives the coevolution of bacteria and phage genomes, through spacer gain or loss and protospacer mutation or deletion, respectively~\cite{82}.  Fundamentally, the phage exposure drives the CRISPR locus to rapidly evolve.  Study of the genomics has indicated that CRISPR evolution is much faster than accumulation of typical nucleotide polymorphisms in bacteria~\cite{56}, and mathematical models of this coevolution have been constructed~\cite{7,101,48}.  These models describe the acquisition and heritability of CRISPR-Cas immune system and characterize this example of Lamarckian inheritance, \emph{i.e.,} where the organism passes on traits acquired during its lifetime to its offspring~\cite{7}.  One analysis paired population dynamic experiments and DNA sequence analysis with a mathematical model of bacteria and phage coevolution~\cite{101}.  Random protospacer mutation brought to light the arms race that occurs between CRISPR-immune hosts and CRISPR-escape mutant phage.  In different parameter regimes, CRISPR-Cas allowed the bacteria to become established and to either extinguish or coexist with phage.  The experiments showed that a high rate of mutation in phage required CRISPR-immune hosts to acquire multiple spacers for complete resistance.  In the language of the Lotka-Volterra predator-prey model, pseudo-chaotic oscillations can occur in the coevolution of CRISPR-immune bacteria and phage~\cite{48}.  Tuning of the phage reproduction leads to stable population equilibria, small periodic oscillations, or pseudo-chaotic oscillation regimes.  This behavior was due to the presence of three population types: CRISPR-immune hosts, sensitive hosts, and phage that had the possibility of acquiring escape mutations. The bacteria's non-linear dependence on the phage population size, and the imbalance between immunity decay and acquisition rates also contributed to the emergence of these three regimes. The pseudo-chaotic regime appeared to capture the heritability and evolutionary instability of CRISPR-Cas loci.

Spatial heterogeneity in the bacteria's surroundings and the phage density have been considered~\cite{87}.  The population densities of uninfected $x$, infected $y$, and resistant $z$ bacteria are
\begin{align}
\frac{\textrm{d}x}{\textrm{d}t} &=  n_{b}x(1-x-y-z)-rn_{v}xy+[1-a(\alpha{})]n_{b}xz-\gamma{}n_{b}x+a(\alpha{})\gamma{}'n_{b}z  \\
\frac{\textrm{d}y}{\textrm{d}t} &=  rn_{v}y(x+(1-\eta{})z)-ry  \\
\frac{\textrm{d}z}{\textrm{d}t} &=  a(\alpha{})n_{b}z(1-x-y-z)-rn_{v}(1-\eta{})zy+[a(\alpha{})-1]n_{b}zx-\gamma{}'a(\alpha{})n_{b}z+\gamma{}n_{b}x .
\label{eq:spatial}
\end{align}
where $r$ is the phage reproduction rate, uninfected bacteria acquire CRISPR spacers to become resistant bacteria at a rate $\gamma$, and resistant bacteria can lose spacers, therefore losing resistance, at a rate $\gamma{}'$.  The uninfected and resistant bacteria populations have a growth rate $a(\alpha{})$ dependent on the cost $\alpha$ of having the CRISPR immune protection, and $\eta$ characterizes the bacterial immunity.  Dependent on the medium, $n_{b}$ is the number of nearest neighbors that the bacteria can access, and $n_{v}$ is the number of neighboring sites phage can access after they burst from an infected bacterium.  The amount of spacer diversity that allows a fast, localized CRISPR response was determined.  The spatial growth of a single bacterial strain was tracked and multiple distinct phage species were followed on a series of lattice sites. How the evolution depended on phage diversity, burst size, phage mutation, diffusion, and latency was explored.  In a well-mixed environment, CRISPR proved to be inefficient in acquiring the needed spacer for a given attack situation.  The system tended toward extreme values of immunity, with a bacterial survival probability of 0 or 1.  In a spatially heterogeneous system, where phage and bacteria are spread in space, the system tended toward intermediate spacer levels.  There were neighborhoods of phage populations and neighborhoods of bacteria populations.  Bacteria with similar spacer numbers clustered together, and phage clustered near bacteria with weaker immunity.

\subsection{Strain divergence}

CRISPR array evolution leads to individuality within an otherwise nearly clonal bacterial population~\cite{61,56}.  Selective pressure from rapidly changing phage populations induces rapid individual-level CRISPR diversification to maintain bacterial population immunity~\cite{56,62,57}, and genomic data analyses have shown that no two sampled strains share the same CRISPR locus~\cite{56}.  A study of \emph{Leptospirillum} group III microbial communities in biofilms collected from Richmond Mine in Redding, CA showed CRISPR loci capable of evolution and modulation of resistance levels on the timescale of months \cite{62}.  In another study, it was found that \emph{S. thermophilus} interactions with phage over just a one-week period led to a genetically diverse population of bacteria \cite{57}.  In this particular experiment, all surviving bacteria had acquired at least one spacer against the phage, and there were multiple subdominant strain lineages.  High spacer diversity within the bacterial population is selected for, since it increases the overall fitness of the population~\cite{196}.

A number of metagenomic studies have been conducted of prokaryote and phage coevolution in natural environments and of the effect that this coevolution has on CRISPR locus diversity.  A high diversity of phage strains was found to lead to a high diversity of CRISPR sequences.  In a three-year study of phage and CRISPR-containing microbial populations in Lake Tyrrell, it was found that archaeal and bacterial populations were overall more stable than their phage counterparts, however there was significant change in the relative abundance and presence of different archaeal strains over time and space \cite{65}.  In a study of hot spring population dynamics of \emph{S. islandicus} archaea from the Nutnovsky Volano region of Kamchatka, 
Russia~\cite{84}.  While it was found that one dominant host genotype coexisted with rare recombinant types, CRISPR analysis reported an even distribution of resistance genotypes within this population.  This is due to rapid evolution of the CRISPR locus relative to the rest of the genome.  Virus-host interactions drove host diversity.  Model predictions and metagenomic data from the Richmond Mine, CA suggested CRISPR's immune memory makes it suited for environments in which viruses persist for long periods or continually immigrate~\cite{64}.

A novel year-long analysis of oral streptococcal in 4 human subjects characterized the CRISPR spacer diversity~\cite{61}.  Streptococcal CRISPR sequences from human salivary microbiome samples were analyzed periodically over 11 to 17 months. Throughout the entire study, 7-22\% of the CRISPR spacers remained constant.  A further 15\%-75\% of spacers were detected only at single time points.  There was a high variation in relative abundance of streptococcal species over time, depending on subject. Interestingly, streptococcal community composition was related to spacer diversity in some subjects.  For example, one subject did not have a dominant \emph{Streptococcus}, but had the highest CRISPR spacer diversity.  There was a high spacer diversity between different subjects, with only 2\% shared between subjects, suggesting that each person was exposed to different virus populations.

\begin{figure}[ht!]
\begin{center}
\includegraphics[width=0.5\textwidth]{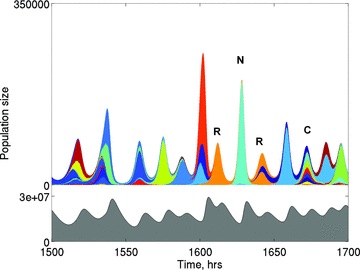}
\caption{Model results showing prominence of diverse bacteria species, each a different color, as a result of CRISPR-Cas targeting of viruses (total population in gray).  `N' denotes a rapidly appearing, novel strain, `C' signifies a time when multiple hosts emerged as coalitions, and `R' identifies a recurring strain. Reused with permission from~\cite{86}.}
\label{fig:diversification}
\end{center}
\end{figure}

A multiscale model of CRISPR-induced coevolution of bacteria and phage was used to study both the strain diversification and population growth~\cite{86}.  The model incorporated ecological events, in which the bacteria and phage growth and decay rates were linked; molecular events, in which sequence matches between CRISPR spacers and protospacers lead to bacterial immunity; and evolutionary events, in which bacteria acquired new spacers and phage acquired escape mutations.  The populations were modeled as bacterial density $x$ of strain $i$ that had a set of spacers $S_i$ and viral density $v$ of strain $j$ that had a set of protospacers $P_j$.  The dynamical equations that govern these densities are
\begin{align}
\frac{\textrm{d}x_i}{\textrm{d}t} &= a_{i}x_{i}\Bigg(1-\frac{\sum\limits_{i}x_i}{q}\Bigg)-(1-r)\sum\limits_{j}[1-M(S_{i},P_{j})]\beta_{ij}x_{i}v_{j} - \gamma{}\sum\limits_{j}M(S_{i},P_{j})\beta_{ij}x_{i}v_{j} \\
\frac{\textrm{d}v_j}{\textrm{d}t} &= (1-r)\rho{}\sum\limits_{i}[1-M(S_{i},P_{j})]\beta_{ij}x_{i}v_{j} + \gamma{}\rho{}\sum\limits_{i}M(S_{i},P_{j})-\sum\limits_{i}\beta_{ij}x_{i}v_{j}-dv_{j},
\label{eq:multiscale}
\end{align}
where $a$ is the bacteria reproduction rate with a carrying capacity of $q$, $\beta$ is the interaction rate between bacteria and virus strains, $\rho$ is the virus burst size, and $d$ is non-CRISPR deactivation of viruses.  $M(S_{i},P_{j})$ is 1 when the bacteria locus contains at least one spacer that matches at least one of the virus protospacers, otherwise it is 0.  If there is a matching spacer and protospacer, $(1-\gamma)$ is the probability of host immunity through CRISPR interference, whereas $(1-r)$ is the probability of bacteria lysis in the absence of a CRISPR spacer.  Starting from communities with low diversity, Figure~\ref{fig:diversification} shows how a high dissimilarity between the coexisting strains could evolve at long times.  Different bacterial strains were able to achieve equivalent levels of resistance via uptake of multiple, distinct protospacers from the phage population.

While CRISPR spacers are quite diverse and dynamic, there is a high conservation of Cas proteins and CRISPR repeats among bacteria in the same genera.  An early study showed that the Cas interference proteins were highly conserved across the two genotypic groups of \emph{Leptospirillum} group II bacteria  \cite{59}.  There was a strong relationship between ecology and genotype gene content, gene sequence, and protein abundance levels of closely related bacteria. In another study, \emph{Synechococcus} bacteria isolated from Yellowstone National Park hot springs were sequenced, and while the microbial strains had highly diverse spacer sequences, they all had similar CRISPR repeats~\cite{85}.

Studies have also investigated the role of the diversity of CRISPR components among loci within a single prokaryotic genome.  \emph{S. thermophilus} has three CRISPR loci, each of which with its own set of \emph{cas} genes and repeats~\cite{33}. Repeat orientation was aligned with \emph{cas} gene orientation.  The first and third loci did have similar sequence architecture, however the second, inactive locus had a degenerate set of repeat sequences.  Most archaea strains have more than one CRISPR-Cas system in their genomes, and these individual CRISPR loci typically do not interact with each other~\cite{124}.  An algorithm had been developed to identify entire CRISPR loci from metagenomic datasets, without the need for prior knowledge of the loci~\cite{167}.  Spacer array reconstruction was reasonable, however it was more difficult to identify spacers in CRISPR loci that did not conserve repeat sequences.  Interestingly, nearly all 43 repeats in one of the CRISPR loci in \emph{Streptococcus sanguinis} locus are different~\cite{78}.
 
\subsection{Selection pressure for survival of the cell}
	
The evolutionary plasticity of bacterial genomes reflects a balance between maintenance of genome stability and tolerance of instability~\cite{132}.  The CRISPR-Cas system brings genome variability but also controls stability by restricting incorporation of mobile elements.  There is a significant fitness cost for a CRISPR system targeting even non-essential host genes~\cite{204}, so these self-targeting spacers tend to be unstable~\cite{166}.  The avoidance mechanisms discovered through engineered spacer experiments were absence of or mutations in the PAM~\cite{163,166}, mutations in the repeats flanking the self-targeting spacer~\cite{163,166}, mutations in the \emph{cas} operon~\cite{163,166}, loss of the self-targeting spacer~\cite{204}, or loss of the self-sequence being targeted~\cite{176,163}.  In one experiment, an artificial mini-CRISPR locus was introduced into a viral genome, and this virus-encoded CRISPR locus was then incorporated into \emph{Sulfolobus solfataricus} bacteria~\cite{204}.  Even when the CRISPR contained spacers targeting a non-essential bacteria gene, recombination with the host CRISPR locus was triggered and the spacer was removed.  The viral CRISPR locus remained intact when it did not contain spacers targeting the host genome.  Another genomic study of the CRISPR system characterized the diversity of Type II and Type IV systems within \emph{E. coli}~\cite{176}.  Self-interference caused degeneration of the Type IV CRISPR-Cas system in some \emph{E. coli} ancestors that were shown to contain a Type II system with a spacer that matched Type IV \emph{cas} sequences.  Strong selective pressure from self-targeting of specific chromosome regions resulted in bacterial genome evolution in the \emph{Pectobacterium atrosepticum} Type I-F CRISPR-Cas system~\cite{163}.

Strong selective pressure for genes that confer virulence or antibiotic resistance leads to the loss of CRISPR function~\cite{102}, loss of the targeted spacer~\cite{102,74}, or loss of the CRISPR system~\cite{105,102,74}.  In \emph{S. pneumoniae} the CRISPR mechanism was shown to block HGT and to be lost under strong selective pressure for virulence or antibiotic resistance~\cite{105}.  The low frequency of bacteria that successfully infected mice in an \emph{in vivo} experiment with \emph{S. pneumoniae} had acquired the gene after losing their CRISPR system.  An experiment with \emph{S. epidermidis} that contained a CRISPR spacer targeting a beneficial plasmid showed that plasmid transfer into the host could occur if the plasmid mutated, the CRISPR spacer was lost, the CRISPR was deactivated or deleted, or the CRISPR response was subdued by other mechanisms~\cite{102}.  Upon being challenged by protospacers that match spacers in their active CRISPR loci and which were associated with essential functions, \emph{Sulfolobus} cells adapted primarily by losing the matching spacer~\cite{74}.  It depended on the species, as \emph{S. solfataricus} averaged large deletions, while \emph{S. islandicus} had a high incidence of specific deletions of single matching spacers by an unknown mechanism. It was suggested that a low level of spontaneous recombination activity occurred to form viable transformants carrying vector-borne protospacers in those cells that deleted their matching CRISPR spacers.	
	
\subsection{Impact of effectiveness}

The abundance of the CRISPR-Cas system in a prokaryotic population is influenced by its effectiveness in conveying immunity.  If CRISPR is more effective, than it is more active and prevalent~\cite{138,136}.  Intriguingly, there was also experimental evidence of a possible positive feedback loop between active spacers that are affording effective protection in a locus and newly acquired spacers~\cite{138}.  All newly acquired spacers of an individual Type I-E  \emph{E. coli} targeted the same strand of the plasmid, suggesting interplay between the interference and adaptation machinery.  This feedback for acquiring more spacers on the same strand as spacers that are already effective was not observed for Type II \emph{S. thermophilus}, suggesting acquisition and interference by Cas9 are not coupled.  Multiple active spacers against different protospacers from the same phage reduced the chance that the phage can evade immunity by point mutation in the PAM or seed region.  In another study, CRISPR was found to be more abundant in hyperthermophilic microbes due to generally lower rates of substitution for phages in thermal habitats~\cite{136}.  Indeed, CRISPR-Cas prevalence is more correlated with thermophilic environments than with simple archaeal taxonomy.

On the other hand, it has been observed that bacteria switch from CRISPR-Cas to a constitutive immune mechanism when high levels of naive bacteria enter an already coevolving host-parasite population~\cite{149,196}.  High levels of \emph{P. aeruginosa} bacterial immigration caused an increase in the frequency of infections.  As the frequency of infection increased, CRISPR protection decreased, which meant that surface modification became the less costly defense.  Bacteria therefore switched from using CRISPR-Cas to a surface modification-mediated defense as the frequency of immigration increased.  Against conjugative plasmids, the intensity of selection favoring CRISPR is weak with very narrow conditions for it to be advantageous~\cite{90}. A mathematical model showed that populations with CRISPR were eliminated when plasmid conferred a growth rate advantage to the infected host, such as antibiotic resistance~\cite{130}.

If there are a large number of possible protospacers~\cite{54} and if CRISPR organizes its spacers well~\cite{130}, CRISPR will be more effective, and therefore more abundant.  Indeed, a mathematical model showed that CRISPR-Cas efficacy increases rapidly with number of protospacers per viral genome~\cite{54}.  Another theoretical model showed that an adaptive immune system may carry a substantial number of receptors for rare antigens, at the expense of receptors for common infections \cite{130}. Experimentally, it has been found that archaeal hosts attempt to balance protecting themselves against persistent, low-abundance viruses and highly abundant viruses that could destroy the host community~\cite{65}.

CRISPR is more prevalent when there is a high viral density or diversity~\cite{90,153}.  Experiments have also shown that the rate of spacer acquisition from phage is proportional to the quantity of these phage in the immediate environment~\cite{153}.  The regulation of CRISPR-Cas mechanisms based on the cost of carrying this type of immune system is discussed in more detail in the next section.  Briefly, mathematical modeling of \emph{E. coli} has shown that a sufficiently high density of phage must persist for the cost of carrying and expressing CRISPR genes to be worthwhile~\cite{90}.  However, CRISPR can be completely lost when the viral diversity is higher than a threshold value, beyond which CRISPR is ineffective~\cite{54}.  A stochastic, agent-based mathematical model of coevolution of host and virus showed that selection for CRISPR-Cas depended on spacer incorporation efficiency $\gamma$, virus population size $v$, number of protospacers per virus $N_p$, viral mutation rate $\mu$, and the fitness cost $\alpha$ of maintaining the CRISPR-Cas system.  In the case where the CRISPR-associated fitness cost is negligible, the characteristic viral mutation rate $\mu^{*}$ is
\begin{equation}
\mu^{*} \approx \frac{\eta{}L}{cv} \approx \frac{4\eta{}N_{p}\gamma}{v},
\end{equation}
where $c$ is the efficiency of the host's constitutive immune protection, $L$ is the CRISPR locus length, and $\eta$ is a constant that represents the correlation between spacers and protospacers.  If the viral mutation rate is greater than $\mu^{*}$, CRISPR-Cas is ineffective and selected against.  It was suggested that CRISPR becomes ineffective in mesophiles because of larger population sizes.

\section{Cost and regulation of CRISPR Activity}
\label{sec:CostAndRegulationOfCRISPRActivity}

The composition and evolution of an immune system is inevitably constrained by the cost of carrying it.  The main factors that regulate CRISPR-Cas activity are locus length, necessity, specificity, and efficiency.  Locus length is a determining factor for the acquisition, retention, and loss of spacers in CRISPR's limited reserve.  If other immune mechanisms are sufficient to defend the host, and CRISPR-Cas is not necessary, these other immune mechanisms will be favored, and CRISPR-Cas may be turned off or replaced completely.  Specificity of the crRNA controls the balance of affinity to the target and cross-reactivity to escape mutants.  The CRISPR-Cas mechanisms appear to be optimized to conserve energy requirements and to use Cas protein machinery and other resources sustainably.

A general theoretical framework was recently developed to predict the optimal repertoire for an organism's defense system receptors to protect against a given distribution of pathogens, minimizing cost and maximizing effectiveness~\cite{130}.  The cost of having an immune repertoire $M_i$ made up of a distribution of receptors $i$ was defined as
\begin{equation}
\textrm{Cost}(\{M_i\}) = \sum_{j}p_{j}\overline{F_j},
\label{eq:cost}
\end{equation}
where $p_j$ is the probability of being infected by antigen $j$ and $\overline{F_j}$ is the average harm caused by this antigen, which is a function of the probability that an encounter between receptor $i$ and antigen $j$ leads to immune recognition and protection.  The model showed how limited numbers of immune receptors can self-organize to provide protection against highly diverse pathogens. It also demonstrated competitive evolution of these receptors due to environmental antigens.  The authors showed that this type of framework could be applied to CRISPR-Cas, to better understand how these organisms protect against diverse threats by organizing an array of specific spacer-mediated responses.

\subsection{Spacer maintenance considerations}

While spacers are the fundamental building blocks of CRISPR-Cas-mediated immunity, acquiring and maintaining them comes with a price.  Results from an experimental study of the interactions between {\emph{Sulfolobale} archaea and various mixtures of the viruses that typically target them suggested that it may be possible for CRISPR adaptation to be mediated by toxin activity that inhibits cell growth~\cite{189}.  Spacer uptake from challenged viruses strongly retarded the growth of some host cultures, with growth typically recovering in 20 days after spacer acquisition in this particular study.  It was confirmed that this was not due to viral infection, but rather the act of spacer acquisition itself, because isolates taken from the host culture that was actively acquiring spacers continued to exhibit retarded grown for an extended period of time.  These growth retardation dynamics possibly occur to provide an opportunity for host cells to uptake spacers before cell division.  Additionally, a study monitoring \emph{S.\ thermophilus} found that the most effective immunity was achieved when all Cas protein sequences were focused on a single highly effective spacer, as cells with this single spacer were more abundant than cells with additional spacers~\cite{63}.  Cas protein complexes are more spread out across a diversity of target sites when there are multiple transcribed spacers, which could reduce immunity, compared to being concentrated on targeting via a single highly effective spacer.

It has been experimentally observed that the CRISPR locus is unable to indefinitely collect new spacers without some spacer loss~\cite{31}.  Several spacer deletion mechanisms were investigated in a mathematical model, namely deleting the oldest spacer, deleting one of the oldest spacers with increasing probability, and randomly deleting a spacer from anywhere in the locus~\cite{8}.  Due to selection for functional spacers, the results from all mechanisms were similar.  Spacer acquisition increases with an increasing viral diversity, and another mathematical model suggested that the CRISPR locus length will only grow until it hits a threshold, at which time it would collapse to zero~\cite{136}.  Due to limitations on length, the CRISPR is less likely to store spacers for threats it is unlikely to encounter again.  For example, in a five-year metagenomic study of population dynamics and spacer diversity in acid mine drainage biofilms and phage, the absence of spacers targeting a particular phage in some mid-locus spacer blocks was evidence for periods of fluctuating exposure to that phage~\cite{60}.

\subsection{Turning CRISPR on and off}

In at least one CRISPR-containing prokaryote, quorum sensing is used to activate or repress the CRISPR-Cas stages of immunity.  In a study of \emph{Pseudomonas aerunginosa}, higher cell densities induced adaptation, \emph{cas} gene expression, and increased interference~\cite{193}.  At low cell densities when the population has a lower risk of becoming detrimentally infected, few cells acquired new spacers, and \emph{cas3}, which encodes for the interference nuclease, was minimally expressed.  The CRISPR-Cas immune system was seven times more effective in eliminating the targeted plasmid when the cells possessed the capability of quorum sensing.  Furthermore, it was demonstrated that pro- and anti-quorum-sensing compounds could be introduced to induce or repress the CRISPR-Cas mechanisms, opening the door for use of quorum-sensing inhibitors to limit the development of bacterial resistance to phage therapy.

Besides CRISPR-Cas, a range of effective antiphage and antiplasmid mechanisms exist in microbes~\cite{250}.  Mathematical models have predicted the dominance of these other immune mechanisms in the host's defense if CRISPR-Cas proves to be ineffective.  CRISPR emerges only at intermediate levels of the host's innate resistance.  For instance, hosts that are already fully resistant via non-CRISPR mechanisms, such as envelope resistance that interferes with phage attachment to a bacteria cell through receptor modification, create narrow conditions for CRISPR to be advantageous~\cite{90}.  If the host survives two-thirds of its predator encounters without the help of CRISPR spacers, CRISPR-Cas becomes too costly to maintain~\cite{136}.  The long-term evolution of host populations as a function of pathogen exposure was studied by Mayer and colleagues in a model that compared innate, adaptive, and CRISPR-like immune strategies~\cite{121}.  The number of expected host descendants in subsequent generations was affected by the protection their immune system afforded during pathogen interaction and the cost of maintaining the immune system in the absence of threat.  The lifetime and frequency of presence of a pathogen in a particular generation selected for different types of host immune systems.  A costly innate immune system was selected for those environments with persistent pathogens, whereas adaptive, non-heritable immunity was best for transient, rare pathogens.  A CRISPR-like immunity that is adaptable and heritable is most advantageous against long-lasting but intermittent pathogens.
	
Westra and colleagues developed a theoretical model with experimental validation of how different ecological conditions drive the selection of infection-induced CRISPR-Cas or constitutive surface receptor modification-mediated immunity~\cite{197}.  The population densities of uninfected $x$ and infected bacteria $y$ that have both constitutive $c$ and induced CRISPR $\gamma$ immune protection rates, and infectious pathogen population density $v$ are governed by
\begin{align}
\frac{\textrm{d}x}{\textrm{d}t} &= [a(c) - q(x+y)](x + fy) - bx - (\beta{}-c)xv + \gamma{}y  \\
\frac{\textrm{d}y}{\textrm{d}t} &= (\beta{}-c)xv - [r + b + \gamma{} + \alpha{}(\gamma)]y \\
\frac{\textrm{d}v}{\textrm{d}t} &= \rho{}ry - dv - (\beta{}-c)xv,
\label{eq:immunemechanisms}
\end{align}
The uninfected bacteria population has a growth rate $a(c)$ dependent on the cost $c$ of having the constitutive immune protection and reduced further based on a sterilization factor $f$ from infected cells $y$, and it decreases due a crowding factor $q$ and death rate $b$.  The $(\beta{}-c)xv$ term represents pathogen transmission to bacteria, based on a constant infection probability $\beta$ and the probability that constitutive protection is successful.  Pathogen virulence factor $r$ is what determines the rate that infected bacteria die and the rate that the pathogen population grows with burst size $\rho$.  The population of infectious pathogens also decreases due to a deactivation rate $d$.  Here, CRISPR protection $\gamma$ is only activated by infection, and it incurs an immunopathological cost $\alpha{}(\gamma)$ on the infected $y$ bacteria.  The impact of the availability of resources and parasite exposure was investigated using this model and in experiments with phage-challenged \emph{P. aeruginosa}.  Since CRISPR-Cas activity was associated with a reduced rate of host replication, high resource environments that led to more infections selected for the host's constitutive defense, whereas resource-limited conditions selected for CRISPR-Cas.  Since surface receptor modification reduced the fitness of the bacteria in the absence of threat, CRISPR-Cas dominated in low-parasite conditions.

A synergy can exist between CRISPR-Cas and other immune mechanisms of the host, as was found in an experimental study of \emph{S. thermophilus} cells that had both an active CRISPR-Ca locus and an active restriction-modification system~\cite{203}.  During restriction-modification, foreign DNA is cleaved at specific recognition sites, and self and non-self are distinguished based on the presence of methyl groups in the bacteria's genome.  The two mechanisms were shown to be compatible and reduce phage infection to a higher degree than either of these mechanisms on their own.  Both systems cleaved their respective target sites in the phage genome, \emph{i.e.}, restriction-modification cleaved specific non-methylated recognition sites and CRISPR cleaved matching protospacer sequences.  Furthermore, whereas phage with methylated DNA sequences were able to evade restriction-modification immunity, CRISPR-Cas interference of these sequences was unimpaired.

\subsection{Incomplete target recognition}
\label{sec:IncompleteTargetRecognition}

CRISPR-Cas immunity in prokaryotes drives the selection of point mutations and recombination in virus protospacers that allow the virus to escape recognition.  However, some CRISPR-Cas systems have the ability to recognize an invading mutated sequence with an imperfect match between its spacer and the protospacer.  The appropriate Cas proteins will then promptly collect more spacers from this virus in order to regain immunity to it, in a process termed `primed' acquisition~\cite{96}.  CRISPR-Cas acquisition of spacers from a new threat is distinguished as `naive' acquisition.  The concept of bacteria regaining immunity through priming has also been integrated into mathematical models of the coevolutionary arms race, where the primed-acquisition positive-feedback loop reduces the ability of an invader to escape via protospacer point mutations~\cite{7}.  Primed acquisition has been observed in multiple experimental studies, and it has been hypothesized that the Cas effector protein:crRNA complex slides along the target DNA, randomly stops at PAM sequences, and recruits more spacers from the same strand~\cite{92,138}.  These studies provide evidence of how encounters with mutants that have tried to evade interference interact with and regulate the CRISPR-Cas response.

The behavior of CRISPR-Cas in  \emph{E. coli}  when encountering foreign DNA sequences that did not perfectly match the bacteria's spacer sequence was studied~\cite{98}.  Point substitutions in the PAM or protospacer strongly decreased the affinity of Cascade:crRNA complex to its target DNA, and instead of sparking its defensive mechanism to cleave its target, the CRISPR inserted new spacers from other PAM-specified locations in the invader's DNA.  The observed primed acquisition mechanism required Cascade, Cas3, and Cas1:Cas2.  However, naive acquisition was observed independently of Cascade and Cas3.  The recruitment of auxiliary genomic stability proteins for spacer acquisition depended on whether the CRISPR was engaged in naive or primed acquisition~\cite{195}.  It was shown that during target surveillance when Cascade bound to invading DNA, Cascade blocked the DNA replication forks by forming an R-loop between the crRNA and protospacer, and RecG dissipated the R-loops to expose the DNA for primed spacer capture.  See Figure~\ref{fig:priming}(A).  However, during native adaptation when Cas1:Cas2 bound to and nicked forked DNA within single strand gaps to collapse replication forks, RecBCD arrived to target these collapsed forks, cut DNA ends, and generated a DNA substrate for spacer capture.  See Figure~\ref{fig:priming}(B).  Both types of adaptation required DNA polymerase I, which appeared to fill DNA gaps by catalyzing new CRISPR repeats during spacer integration.	

\begin{figure}[ht!]
\begin{center}
\includegraphics[width=0.7\textwidth]{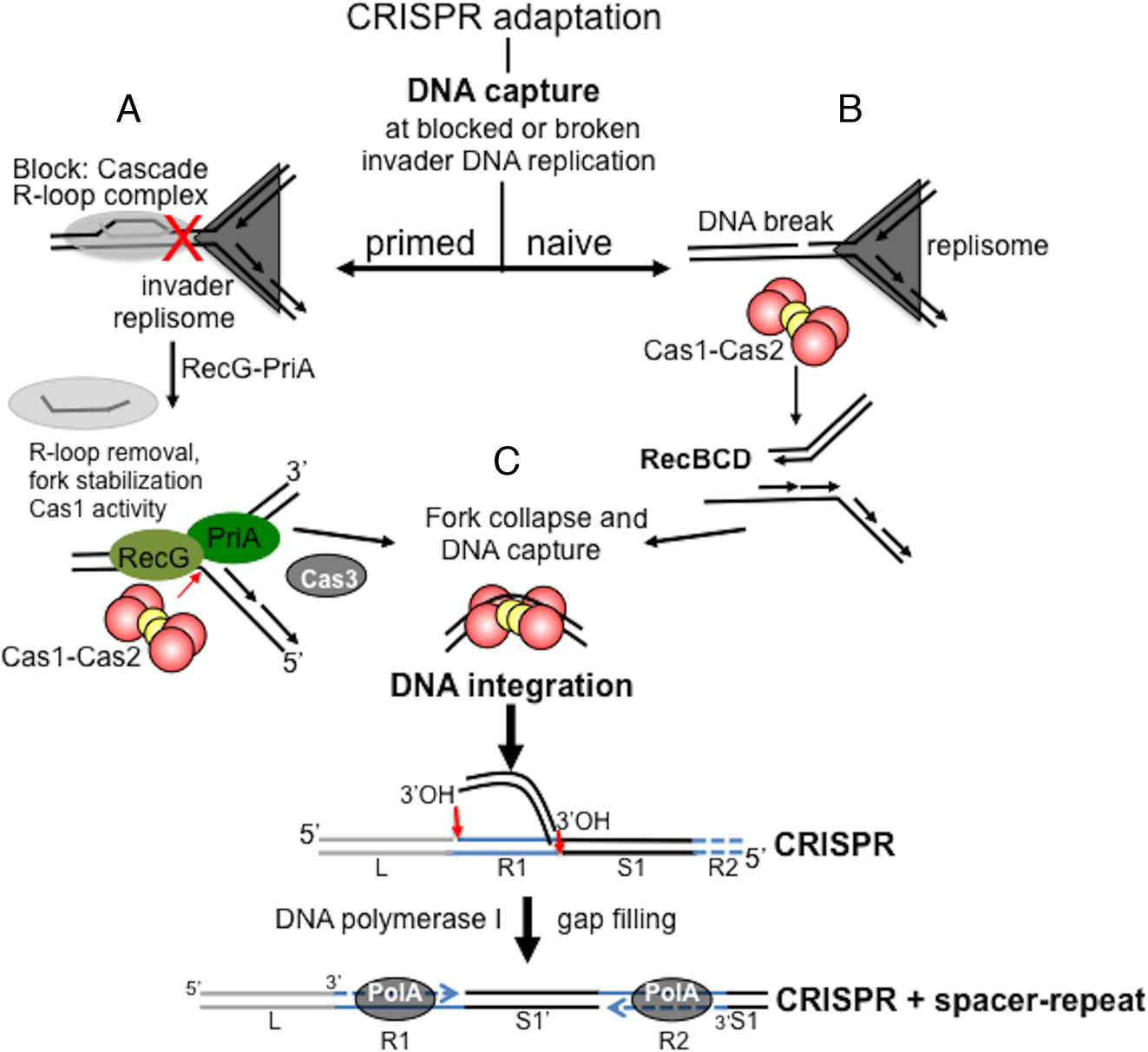}
\caption{The two pathways of spacer acquisition in  \emph{E. coli}  are putatively regulated by the presence of Cascade.  (A) If Cascade binds to invading DNA without cleaving it due to an imperfect but tolerable match, this sparks primed acquisition of the sequence via Cas1:Cas2 and other proteins.  (B) On the other hand, if the CRISPR does not have a matching spacer within the mismatch tolerance threshold, Cascade will not be drawn to the invading DNA.  Instead, naive acquisition is started by Cas1:Cas2.  Reused with permission from~\cite{195}.}
\label{fig:priming}
\end{center}
\end{figure}

An in-depth assessment of Type I-E sequence requirements for interference versus priming revealed five PAMs for the former and 22 PAMs for the latter~\cite{99}.  Cascade and Cas3-mediated interference readily occurred even with up to five mutations at 6-nt interval positions throughout the protospacers and two-three more mutations in the non-seed region.  Primed acquisition occurred for targets with up to 13 mutations throughout the PAM and protospacer regions that had escaped interference.  It was suggested that priming may explain the selection to retain old, imperfect spacers in the CRISPR locus, since they are still useful for priming from mutated or related invaders.  Fluorescence resonance energy transfer (FRET) microscopy was used to demonstrate that the Cascade:targetDNA conformation depends on the presence of mutations in the PAM and seed regions, and this conformation dictates interference or primed adaptation activity~\cite{190}.  As shown in Figure~\ref{fig:priming2}, during target DNA binding, the large Cascade subunit Cas8e can either have a `closed' or `open' conformation, prompted by mutations in the protospacer PAM, protospacer seed, or a particular motif in Cas8e, `L1.'  Cas3 has been observed to cleave invading DNA into spacer-length pieces of 30-100 nt with PAM sequences on the 3' ends, and Cas1:Cas2 appears to then recycle these DNA degradation products to form new spacers in the CRISPR locus~\cite{179}.  When the original spacer triggers sufficiently strong interference or when Cas3 activity is very high, priming acquisition does not occur.

\begin{figure}[ht!]
\begin{center}
\includegraphics[width=0.7\textwidth]{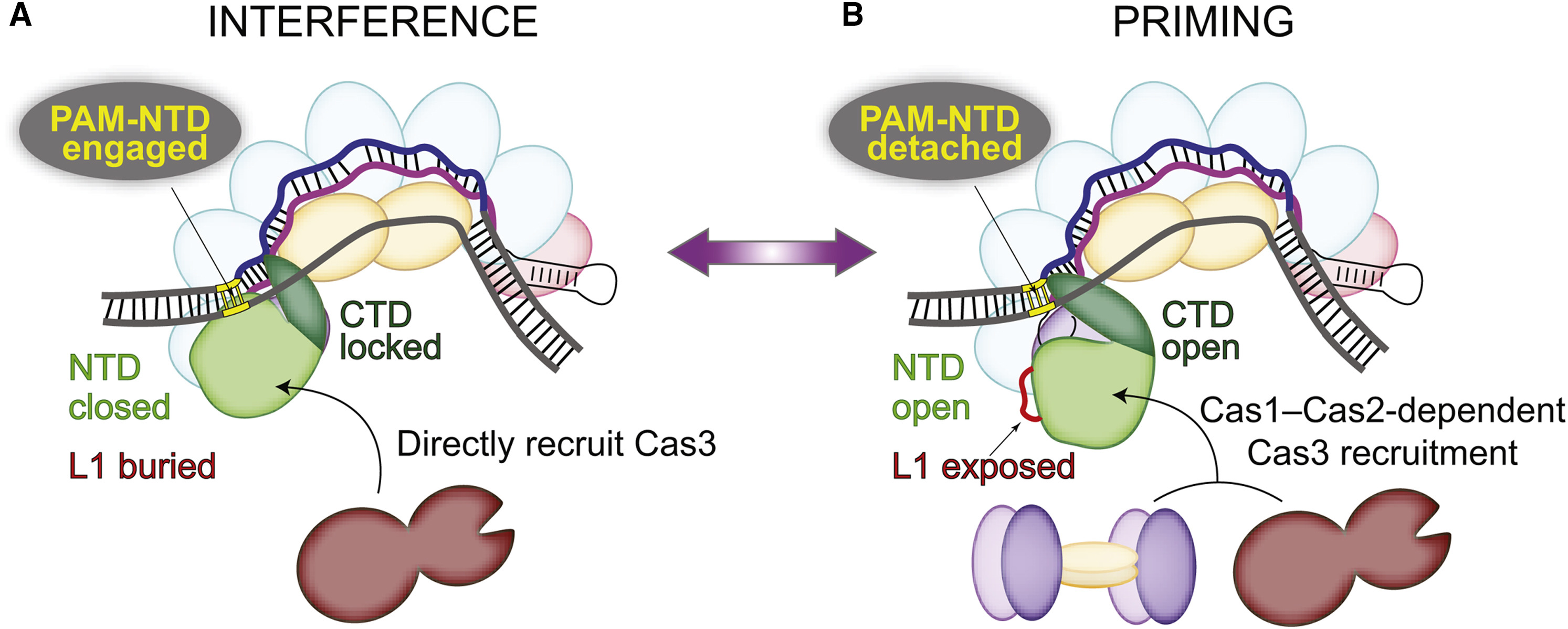}
\caption{The conformation of Cascade during target binding determines if the interference or priming pathway will be activated.  (A) For the closed conformation, the C-terminal domain is locked, the N-terminal domain is closed and engaged with the PAM, and the L1 motif is buried, which leads to recruitment of Cas3 and target DNA interference.  (B) For the open conformation, the C- and N-terminal domains are both open, and the L1 motif is exposed, which sparks primed adaptation from recruitment of Cas1:Cas2 and Cas3.    Reused with permission from~\cite{190}.}
\label{fig:priming2}
\end{center}
\end{figure}

Kiani and colleagues developed a programmable, multifunctional Cas9:sgRNA system that takes advantage of how CRISPR-Cas activity can be regulated by the extent to which a target has been recognized~\cite{188}.  Typically the Cas9 will bind and cleave a target specified by the sgRNA to perform genomic editing, while a deactivated Cas9 is engineered to bind the target without cleaving it to perform gene expression control.  Instead of having to engineer two separate systems, the sgRNA length was altered to dictate Cas9 nuclease activity for either genomic editing or gene expression control at different target sites within the same cell.  Longer sgRNAs showed typical, robust nuclease cutting activity, while shorter sgRNAs of 16-nt or less showed loss of the Cas9 cutting function.  The innovative, multifunctional system therefore employed both long, 20-nt sgRNAs for binding and cleavage and short, 14-nt sgRNAs for binding and subsequent gene regulation.  By fusing Cas9 to a powerful transcription activator domain, the user gains simultaneous control of RNA production regulation and DNA cleavage.

\subsection{Energy, efficiency, and stability}

Some aspects of CRISPR-Cas appear to be optimized for low energy consumption and efficient use of Cas proteins.  For example, the scanning and recognition process of the Cascade surveillance protein complex does not consume adenosine triphosphate (ATP)~\cite{79}.  Furthermore, the Cascade morphology sequesters every sixth base of crRNA:targetDNA binding~\cite{20,222}, and so there is no topological distortion of the protein if there is a mismatch between the crRNA and target DNA at these positions.  As a result, there is no associated energy cost for sixth basepair mismatches~\cite{20}.  An example of efficiency is in Type V-A CRISPR-Cas systems, which use the same dual-reaction Cas protein for both RNA cleavage during expression and DNA cleavage during interference.  Two distinct motifs were identified on the Cas12a of \emph{Francisella novicida}~\cite{112}.  The endoribonuclease motif was specific to ribose for processing pre-crRNA into crRNA and could not cleave DNA, while the endonuclease motif only cleaved target single-stranded DNA and double-stranded DNA and used the crRNA produced in the first reaction as its guide.

Protospacers with frayed nucleotide ends appear to be preferentially acquired~\cite{115}.  The frayed nucleotide end of protospacers is presumably preferred because it requires lower free energy for Cas1:Cas2 to bind to protospacers for spacer acquisition.  The terminal nucleophilic 3'-OH of each protospacer strand needs to enter a constrained channel that leads to the active sites of Cas1.  X-Ray crystal structures revealed that Cas1:Cas2 complexes therefore prefer protospacers with five overhanging 3' nucleotides, instead of completely double-stranded 33-bp protospacers, single-stranded DNA, or substrates with 5' overhangs.  A lower free energy is required for Cas1:Cas2 to bind to these substrates compared to perfectly duplexed ends, which would need to be splayed prior to capture.

Since acquisition depends on the Cas9:sgRNA complex and since RNA can have a limited lifetime \emph{in vivo}, stability can be a concern in applications of CRISPR-Cas.  The stability of engineered sgRNA was highly dependent on being in complex with a Cas9 protein and on the length of the sgRNA, with shorter guides being less stable~\cite{245}.  This stability is an important factor to consider when trying to implement Cas9:sgRNA systems for \emph{in vivo} editing.  It was observed that the ribonucleoprotein had a much longer residence time when in contact with a perfectly matching sequence.  The maximum three-hour dwell time decreased to as low as two minutes if there were considerable mismatches. The shorter dwell time on imperfect matches was also correlated with lower CRISPR cleavage activity.

\section{Non-immunological mechanisms}
\label{sec:NonImmunologicalMechanisms}

The CRISPR-Cas system seems to be more than just a means for providing immunity to its host through interference of infection.  It plays a role in maintaining genome integrity, acquiring new genetic material to adapt, and controlling transcription~\cite{23}.  These functions were suggested by studies showing that spacers in both lactic acid bacteria and archaea include about 20\% matches to self-chromosomes~\cite{163}.  Self-targeting spacers can cause autoimmunity, but it is now thought that they may also have a regulatory or abortive infection role~\cite{156}.  In some pathogenic prokaryotes, CRISPR appears to increase virulence and evasion of the pro-inflammatory response of their host, leading to a higher probability of successful infection~\cite{23,156}.

In 2012, CRISPR was harnessed for genetic engineering when Jinek and colleagues identified the dual-RNA structure responsible for directing Cas9 to cleave a particular DNA target and subsequently engineered a single RNA chimera to successfully perform the same function on specified DNA targets~\cite{36}.  Since then, the Cas9 structure, assembly with the sgRNA, and molecular mechanisms of target search and cleavage have all been heavily studied~\cite{212}.  Owing to its genetic precision and single guide assembly, the use of CRISPR-Cas-based technology has become the preferred method of genome editing and exogenous transcription control (Figure \ref{fig:biotech}).  Further work is underway to mitigate some of the limitations, which include having to match the PAM sequences of the Cas9 species being used, preventing off-target mutagenesis, and making high efficiency sgRNAs.  

\begin{figure}[ht!]
\begin{center}
\includegraphics[width=0.9\textwidth]{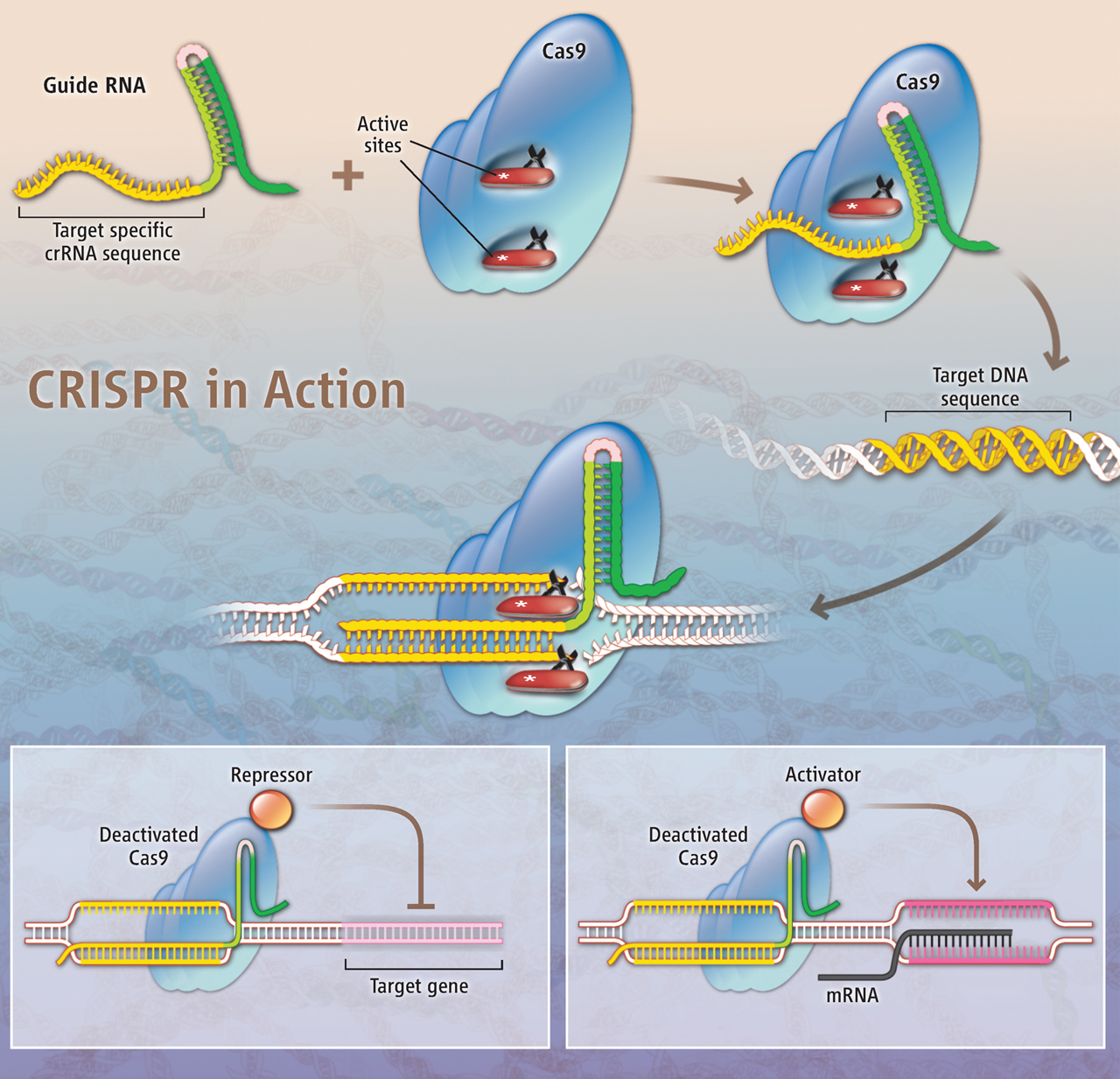}
\caption{CRIAPR-Cas9-based technologies are being use for sequence-specific genome engineering.  (Top) The sgRNA, made up of a crRNA (yellow) and stabilizing tracrRNA (green), in complex with Cas9 binds to a target sequence and performs exact double-strand DNA cleavage.  (Bottom) If the Cas9's cleavage sites are deactivated, the Cas9:sgRNA complex can be used to regulate inhibition or expression of a target gene, by inclusion of a repressor or activator to the Cas9 protein. Reused with permission from~\cite{D}.}
\label{fig:biotech}
\end{center}
\end{figure}
	
\subsection{Endogenous genomic editing}

Since the majority of CRISPR spacers target mobile genetic elements~\cite{171,166}, and self-targeting spacers are not usually evolutionarily conserved, self-targeting spacers initially appeared to be just an `Achille's heel' of the CRISPR-Cas system~\cite{166}.  However, it is not uncommon for small RNAs to be used in gene regulation, and specifically gene silencing, through the inhibition or degradation of messenger RNA~\cite{49}.  RNA interference in eukaryotes helps to prevent the propagation of DNA that does not specifically contribute to the cell's reproductive success.  RNA interference has other roles in genome maintenance and repair.  The similarities between RNA interference and CRISPR-Cas have led researchers to believe self-targeting CRISPR spacers may analogously function as a gene regulation system for endogenous transcription control and genome homeostasis~\cite{23}.

A quarter of the \emph{Streptococcus agalactiae} genome is interspersed with genomic islands formed by integrative and conjugative elements that had been passively propagated during chromosomal replication and cell division, and it is now believed that spacers are likely to have controlled the diversity of mobile genetic elements in these  strains~\cite{171}.  In experiments with the potato phytopathogen \emph{Pectobacterium atrosepticum}, large scale genomic changes were demonstrated to occur as a result of self-targeting CRISPR spacers~\cite{163,233}.  See Figure~\ref{fig:endogenousediting}.  This bacteria was engineered to self-target a chromosomal gene within a horizontally acquired pathogenicity island, though the genome naturally contains this self-targeting spacer with a single PAM mutation~\cite{163}.  With crRNAs guiding host chromosome cleavage, most cells could not readily recover, but a small subpopulation survived with morphological changes, \emph{e.g.}, elongation and filamentation.  On the other hand, the surviving healthy population had either excised or modified the targeted pathogenicity islands.  It appears that self-targeting contributes to bacterial fitness and genome mosaicism via selection for the deletion of islands or other parts of the genome~\cite{233}.

\begin{figure}[ht!]
\begin{center}
\includegraphics[width=0.9\textwidth]{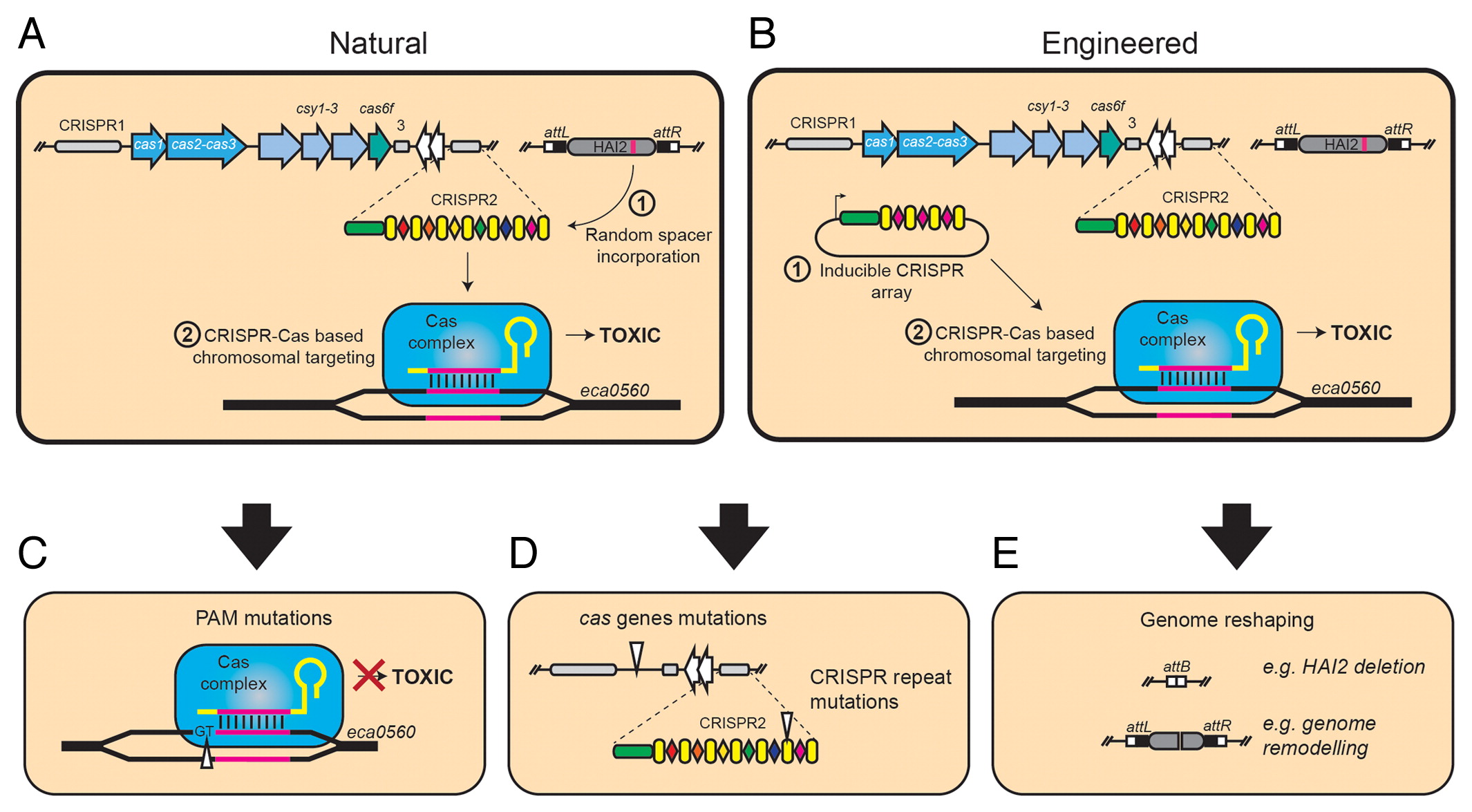}
\caption{(A) Natural self-targeting spacers are toxic to the host cell, (B) but experiments that have purposely engineered self-targeting spacers found various escape routes to prevent cell death.  These include (C) mismatches between the Cas protein's PAM and the host genome's adjacent sequence motif, (D) mutation of the Cas genes or repeats that disrupt CRISPR-Cas function, or (E) removal or recombination of the targeted sequence from the host genome.  Reused with permission from~\cite{233}.}
\label{fig:endogenousediting}
\end{center}
\end{figure}

\subsection{Increased virulence and abortive infection}

Experiments have shown that CRISPR-Cas can play an important role in boosting virulence and allowing pathogenic organisms to evade host defenses to replicate within the host.  One example is the mysterious dependence of intracellular bacterial growth on Cas2 during amoebae infection~\cite{151}.  A single Type II-B CRISPR locus was found in \emph{Legionella pneumophila} and expressed during the exponential phase growth of the bacteria in all types of media, \emph{i.e.}, nutrient-rich and nutrient-poor.  It was also expressed during intracellular infection of aquatic amoebae \emph{Harmannella} and \emph{Acanthamoeba} and of human macrophage.  Mutants lacking all \emph{cas} genes grew normally in the different media and during infection of macrophage, however during infection of amoebae, mutants lacking \emph{cas2} were significantly impaired.  Cas2 apparently mediates or facilitates this type of infection through a physiological mechanism entirely different than the typical CRISPR-Cas immunity function.  Another bacteria physiological function mediated by Cas proteins is biofilm formation in \emph{P. aeruginosa}, which is important to the pathogenic life cycle of this bacteria~\cite{205}.

\emph{Francisella novicida} uniquely uses its CRISPR for self-genome targeting to evade the immune response of eukaryotic cells that it infects and to resist antibiotics~\cite{155,157}.  \emph{F. novicida} expresses a bacterial lipoprotein that lowers resistance to membrane stressors, such as antibiotics or the host cell pro-inflammatory response~\cite{157}.  In the case of facing a eukaryotic immune response, loss of bacterial envelope integrity was linked to increased inflammasome activation in the eukaryotic host.  A naturally expressed crRNA in the bacteria's CRISPR-Cas system targeted the endogenous transcript that encodes for the above-mentioned lipoprotein~\cite{155}.  Rather than \emph{F. novicida} altering its outer membrane structure or increasing its surface charge, it was its Cas9:crRNA that was proven to be responsible for promoting resistance to membrane damage from stressors.  The transcription level of this lipoprotein gene and the secretion of the host's pro-inflammatory cytokine largely increased when either \emph{cas9} or the special self-targeting crRNA was deleted, confirming the lipoprotein was in fact being controlled by CRISPR-Cas.  During intracellular infection, \emph{cas9} and the self-targeting crRNA were expressed at about the same time.  This CRISPR-Cas system also played a crucial role in the regulation of bacterial physiology and antibiotic resistance~\cite{157}.

Abortive infection is a mechanism of programmed death of an infected cell that occurs to prevent a bacteriophage from further reproducing~\cite{217}.  Though this is normally an independent and complementary immune mechanism to CRISPR-Cas, self-targeting spacers and RNA-targeting systems are potentially additional means by which host cells may program death.  For example, the oral bacterium \emph{Leptotrichia shahii} Type VI-A CRISPR-Cas system appears to target RNA for programmed cell death to abort population infection~\cite{169}.

A coevolutionary model that investigated the role of CRISPR autoimmunity in preventing phage reproduction found that within regimes where CRISPR is advantageous, there were two important defense pathways to combat the phage, interference $\gamma$ and toxic self-targeting $\alpha$~\cite{51}.  The population densities of uninfected bacteria $x$, and infected bacteria $y$, and pathogens $v$ were modeled by
\begin{align}
\frac{\textrm{d}x}{\textrm{d}t} &= ax(1-\frac{x+y}{q}) + \gamma{}s_{yv}y - \beta{}xv - \eta{}\alpha{}s_{xx}x \\
\frac{\textrm{d}y}{\textrm{d}t} &= \beta{}xv - \gamma{}s_{yv}y - \alpha{}s_{yy}y - r{}y \\
\frac{\textrm{d}v}{\textrm{d}t} &= \rho{}ry - dv - \beta{}xv
\label{eq:abortive}
\end{align}
The uninfected bacteria population increases with a growth rate $a$ limited by the environmental carrying capacity $q$ and decreases due to the infection rate $\beta$.  The $\gamma{}s_{yv}y$ term represents successful CRISPR interference from an uninfected bacteria containing a spacer $s$ that matches the phage.  The phage has a lysis rate $r$ with associated burst size $\rho$, and a death rate $d$.  Autoimmune events occurred from self-targeting spacers in both the infected $s_{yy}$ and the uninfected bacteria $s_{xx}$, however the scale factor $\eta$ controlled how much autoimmunity occurred to cells in the uninfected state.  When $\eta{}=0$, autoimmunity never happens outside of an infection, and when $\eta=1$, there is no difference in the rate of autoimmunity between the two cell states.  Interference was the typical mechanism of immunity, in which the CRISPR contained a useful spacer and attacked an invading DNA sequence; toxic self-targeting was activated as an abortive infection mechanism when the CRISPR failed to protect the cell from an invader.  The phage population density was much lower when these two mechanisms acted together, in comparison to interference alone.

\subsection{CRISPR-Cas9-based genetic engineering}

Before harnessing the CRISPR-Cas system, two fairly efficient methods of performing genome editing were phage-mediated recombination and transcription activator-like effectors (TALEs).  For phage-mediated recombination, linear DNA cassettes (30-50 bp) synthesized \emph{in vitro} were introduced through electroporation and precisely recombined \emph{in vivo} for gene replacement in bacteria using the homologous recombination system of a defective prophage~\cite{17}.  TALEs were site-specific DNA-binding proteins from a plant pathogen that were customized to modulate the transcription of specific endogenous genes in human cells, and they required the design and assembly of two nucleases for each target site~\cite{45}.  These previous attempts were unfortunately stunted by difficulties in protein design, synthesis, and validation for specific DNA loci of interest~\cite{3}.  After their invention, CRISPR-Cas-based genome editing technologies quickly became preferred for their minimal targeting site requirements, ease of engineering and delivery into cells, and ability to perform multiplex gene editing with multiple sgRNAs co-transformed at once~\cite{69}.  

Genomic {\bf in}sertions and {\bf del}etion{\bf s} (indels) are performed by the Cas interference proteins, which are programmed with a sgRNA to make specific cuts, and endogenous or exogenous DNA repair systems.  Typically the Cas9 protein derived from \emph{S. pyogenes} is used.  Recently, researchers have also started developing editing systems that repurpose Cas12a~\cite{112,244}.  After cleavage, homology-directed repair (HDR) can be precisely designed with a nearby homology donor to work at the gene scale~\cite{187}.  After the broken chromosome ends are cut out to yield single-stranded DNA tails, they invade a homologous chromosome to copy its genetic info, and then gap-repair DNA synthesis and ligation take place.  Conversely, non-homologous end joining (NHEJ) is error-prone and unpredictable, so it is typically used for small indels or to induce mutations~\cite{18}.  With no homology donor, the NHEJ nuclease cuts out the damaged DNA, the DNA polymerase fills in new DNA, and the ligase restores integrity to the DNA strands with a substantial junctional diversity in repaired outcomes.

When CRISPR-Cas9 was first starting to be incorporated into existing genome editing techniques, it was used for selection against unedited bacterial cells~\cite{16}.  Here, the desired mutation is introduced into a bacterial genome by a transformation template and then a CRISPR-Cas9 cassette, which is programmed to target the original, non-mutated sequence, is added to fatally cleave the wild-type cell genomes~\cite{15}.  The resulting population will contain only the strains that had successfully incorporated the desired mutation.  In this way, CRISPR-Cas9 is especially valuable for efficiently recovering subtle changes that have been introduced.  For instance, after minimum-effort genome editing was performed on the PAM of a gene in \emph{Lactobacillus reuteri} using oligonucleotides and RecT proteins, a CRISPR system was injected into the cells to easily identify and eliminate unedited cells~\cite{164}.

CRISPR-Cas9 has been used in plant breeding to perform gene and whole gene family knockout and to induce genetic variation in crops such as wheat, maize, rice, sorghum, tomato, and orange~\cite{5}.  The first plants genetically modified with this gene editing approach were \emph{Oryza sativa} (rice) and \emph{Triticum aestivum} (wheat)~\cite{68}, though the redundancy of genes in the wheat genome make it more difficult to completely knock out a gene~\cite{69}.  Targeted gene knockout was performed in \emph{Solanum lycopersicum} (tomatoes) and was heritable, however the mutated plants exhibited limited fertility~\cite{70}.  In the Arabidopsis plant, CRISPR-Cas was used to induce one-basepair insertions or short deletions into multiple genes that successfully propagated down through three subsequent generations~\cite{71}.  Additionally, an antibiotic resistance cassette was successfully integrated into this plant with reduced off-target activity due to the use of two Cas9:sgRNAs, each one targeting a single DNA strand, instead of using one Cas9:sgRNA for a double-strand break~\cite{73}. 

This editing technology can now induce precise cleavage at endogenous genomic loci in mouse and human cells~\cite{29}, as well as genetically modify somatic human cells with HDR based on a repair donor~\cite{38}.  Applications to disease therapeutics in animal models and clinical trials are described in Sections~\ref{sec:AnimalModels} and~\ref{HumanDiseaseTherapeutics}.  Heritable germline mutations have been achieved in model organisms, such as in the nematode \emph{Caenorhabditis elegans}~\cite{238} and in the parasitoid jewel wasp \emph{Nasonia vitripennis}~\cite{240}.  In the former case, worms were microinjected with vectors encoding Cas9 and the sgRNA of interest, whereas in the latter, wasp eggs were removed from their fly hosts, injected with Cas9 and sgRNA, and then replaced back into the host.  A balance had to be found between having high enough concentration of Cas9:sgRNA for efficient cleavage, while avoiding toxic, off-target effects.  Both systems showed great potential for generating heritable genomic changes in other multicellular eukaryotes.

\subsection{Exogenous transcription control: CRISPRi and CRISPRa}

A catalytically {\bf d}eactivated {\bf Cas9} (dCas9) can be fused to activators or repressors to encourage or inhibit RNA polymerase binding to desired promoter sequences~\cite{23}.  For most of these epigenetic studies, in which gene expression is controlled by non-genetic means, dCas9 is developed from the \emph{S. pyogenes} Cas9 with silent mutations in the RuvC and HNH nuclease domains to disrupt cleavage.  The use of dCas9 directed by a custom sgRNA is a quick, versatile, and economical method of controlling transcription, since creating a particular guide only takes two short custom oligonucleotides and a cloning step~\cite{110}.

The {\bf i}nhibition of expression of specific genes, known as CRISPRi, can be carried out in one of two ways.  The first is by targeting the coding DNA strand of the protein-coding or untranslated region to block transcription elongation; the second is by targeting either the coding or the transcribed strand of RNA polymerase-binding sites to block transcription initiation~\cite{19}.  Qi and colleagues developed a CRISPRi-dCas9 system, introduced it into  \emph{E. coli}, and, unlike traditional gene knockouts, showed the system was reversible by simply disassociating dCas9 from the target site~\cite{14}.  The system was easily deliverable via natural DNA horizontal transfer~\cite{147}.  Gene silencing is more efficient when the sgRNA is at least 20-25 nt and when there is a small distance between the target and transcription sites~\cite{19}.  The dCas9 can target distal regulatory elements, such as enhancers 10 to 50 kb away from the gene of interest, and it was found to be specific and efficient when bound to repressors such as the Kr{\"u}ppel-associated box~\cite{109}.  CRISPRi is more effective than RNAi at blocking transcription in eukaryotes because CRISPR does not naturally occur and therefore does not interfere with endogenous RNA gene regulation~\cite{3}.

On the other hand, by combining dCas9 with a transcriptional {\bf a}ctivation domain, expression can be increased for endogenous genes according to the sgRNA in a technique known as CRISPRa.  Multiple sgRNAs targeting different genes can function efficiently together within the same mammalian cell~\cite{160}.  CRISPRa has also been used to achieve over-expression of genes in human cells for cell and gene therapy, genetic reprogramming, and regenerative medicine~\cite{110}.  Recently, a flexible CRISPRa system that could be be used with a variety of dCas9 proteins was created using an acetyltransferase activation domain for high-specificity gene regulation at both promoter-proximal and -distal locations~\cite{9}.

Many research groups have utilized the versatility of these dCas9-based systems to perform both CRISPRi and CRISPRa with high specificity and efficiency~\cite{24,143}.  A sgRNA that targets upstream of the transcription start site of the gene of interest will lead to activation, whereas one that targets downstream of the start site will cause gene repression~\cite{141}.  Unique sgRNAs were tested with a high-throughout screen around transcription start sites for about 50 genes, resulting in the creation of genome-scale CRISPRi and CRISPRa libraries with ten sgRNAs for each gene that maximized efficacy and minimized off-target effects~\cite{13}.  As CRISPR-based gene regulation techniques are being pushed towards \emph{in vivo} application in humans, it has been especially important to create these libraries with sgRNA sequences that have maximized efficacy and minimized off-target effects~\cite{13,141}.

CRISPR has been used to process RNA as well.  Rather than use the Cas9 or Cas12a interference machinery as is most commonly done in biotechnology, Qi and colleagues utilized \emph{P. aeruginosa}'s Cas6f, which is the endonuclease that cleaves pre-crRNA into crRNAs during the expression phase~\cite{131}.  They developed a synthetic RNA-processing platform to efficiently and specifically cleave precursor messenger RNA (mRNA) for gene regulation in archaea, bacteria, and eukaryotes.  The cleavage was induced at desired loci by inserting Cas6f's recognition element, which is the 28-nt repeat sequence for this family of Type I-F systems.  After the recent discovery of the RNA-targeting Type VI CRISPR system, Abudayyeh and colleagues made use of the Cas13a1 interference protein from \emph{L. shahii}~\cite{169}.  They engineered a sgRNA to successfully target the single-stranded RNA of specific mRNAs \emph{in vivo}.

\subsection{Inducible systems: iCRISPR}

Gene expression and editing can be precisely controlled non-invasively over space and time by {\bf i}nducing CRISPR-Cas activity via chemical or optical means in a technique sometimes termed iCRISPR.  Chemical control has notably been achieved through doxycycline-induced activation of Cas9 activity~\cite{46,150}.  During iCRISPR genome editing, off-target mutations were limited by using a mutated Cas9 that created only single-strand nicks and two closely spaced sgRNAs to target alternate DNA strands~\cite{46}.  By restricting where and for how long Cas9 is expressed in the organism, tissue-specific gene deletions and reduced toxicity were achieved.  In mice, Cas9 induction was strongest in the intestine, skin, and thymus, but it was also able to be induced in the liver.  Doxycycline-activated expression of dCas9 fused to a repression domain was used to study early cell differentiation and to model disease development~\cite{150}.  iCRISPRi was shown to be highly versatile, adaptable to multiple cell lines, and completely reversible by removing doxycycline.  This technique was especially efficient when targeting near the transcription start site.

Photoinducible activation of Cas9 has been demonstrated for high precision control over genomic editing and both endogenous and exogenous gene expression~\cite{10,11,106}.  In one system, the Cas9:sgRNA crystal structure was studied to determine the optimal split site, and the protein was then engineered to have blue light-activated dimerization domains~\cite{10}.  The Cas9 fragments attached when irradiated to perform indel mutations and then separated and ceased cleavage activity when radiation was turned off.  Similarly, optogenetic transcriptional control was achieved with heterodimerization proteins attached to two dCas9 fragments, showing increased transcription of the target gene in mammalian cells when illuminated by blue light~\cite{11}.  A UV light-activated system used patterned illumination to activate a Cas9, which is otherwise inhibited from being bound to photocaged lysine, for endogenous gene silencing to study a transmembrane receptor associated with leukemia and lymphoma~\cite{106}.

Interestingly, Oakes and colleagues identified an ``allosteric switch'' on Cas9, which allows regulation of the protein's activity by binding an effector molecule to a site other than the protein's active site~\cite{113}.  They searched for potential insertion sites within the distinct Cas9 domains that would not disrupt its RNA-guided DNA binding and cleavage functions. Possible sites were found within the helical recognition lobe, within the linker between the recognition and nuclease lobes, within the HNH domain and RuvC region, and within the PAM-interacting domain.  This ligand-dependent activation of Cas9 worked as a tunable CRISPRi and editing system with proven reversibility and versatility in both prokaryotic and eukaryotic cells.

\section{Applications in biotechnology}
\label{sec:ApplicationsInBiotechnology}

\subsection{High resolution live cell imaging}

Superresolution imaging of chromatin has been improved by fusing a photoactivatable fluorescence protein, such as green fluorescence protein (GFP), to a dCas9 programmed to bind specifically to the site of interest.  For instance, the subdiffraction features of the nucleotide sequences at each end of chromatids, known as telomeres, were observed through this specific labeling~\cite{120}.  The difference in size of the telomeres in different types of mammalian cells was also quantified.  Increased fluorescence signal intensity in another imaging study was achieved by binding an appropriate protein scaffold to dCas9 to recruit multiple copies of GFP to the target site~\cite{144}.  With a brighter signal, a lower power excitation laser can be used and the cells can be imaged for longer without photobleaching.  This method is comparable in specificity and efficiency to fluorescence \emph{in situ} hybridization, without requiring sample fixation and dehydration~\cite{142}.

Live cell imaging with fluorescently tagged dCas9 provides insight into chromosome conformations and dynamics during cell division~\cite{142}.  The telomeres displayed confined movement at timescales shorter than 5 s, and macroscopic diffusion though directional transport at longer timescales.  These observed dynamics were comparable to those measured by time-resolved fluorescence imaging, without perturbing the binding or localization of other proteins.  Furthermore, a flexible, two-component labeling approach has been developed in conjunction with dCas9 to further reduce perturbation, photobleaching, and phototoxicity during live cell imaging~\cite{148}.  Here, dCas9:sgRNA transfection was used to specifically introduce a small peptide, known as an epitope tag, to a gene of interest.  As the peptide did not function on its own, a fluorescent protein unit, which also does not function on its own, was introduced and fluoresced after complementation with the peptide.  This system was both versatile, with the possibility of using a variety of fluorescent protein units, and specific, with CRISPR-mediated gene targeting.

These dCas9-based advances in superresolution microscopy have also been applied to studying the diffusion and chromatin binding of Cas9 as it searches for and cleaves target DNA in mouse cells~\cite{126}.  The \emph{in vivo} occupancy times of dCas9, labeled with a ligand that expresses blue fluorescent protein, were measured to understand the relative kinetics of on- versus off-target binding.  Single-particle tracking was used to visualize how dCas9 explored large eukaryotic genomes, showing that dCas9 demonstrated a diffusion-dominated behavior when encountering off-target sites.

\subsection{Encoding information}

Guernet and colleagues used CRISPR-Cas9 to introduce specific point mutations into tumor cells in order to track clonal dynamics in a large population~\cite{194}.  Complex `barcodes' were created in thousands of cells by using CRISPR-Cas9 to make double-strand breaks at specific genomic locations and using HDR to insert a series of silent point mutations at these locations.  These genetic labels could then be read by realtime quantitative PCR (polymerase chain reaction) to determine the proportion of modified cells within the mass population and to trace the emergence of subpopulations of tumor cells containing the barcode mutations.  This technique was used to show how receptor inhibition therapies could result in the selection of subpopulations with alternative resistance mechanisms, to assess the effects of combined drug therapies, and to evaluate the genomic level effects of repairing oncogenic driver mutations in tumor cells.

Shipman and colleagues have recently exploited the fact that CRISPR-Cas creates an immunological memory to deliberately encode information within bacterial genomes~\cite{133}.  They generated a record of defined DNA sequences in the Type I-E CRISPR-Cas locus of  \emph{E. coli} by directing it to capture synthetic protospacers from specific oligonucleotides \emph{in vivo}.  These protospacers were readily integrated as spacers, however the inclusion of a PAM increased the efficiency of acquisition and caused mostly forward orientation additions.  Shipman and colleagues were able to demonstrate the delivery of their specified DNA sequences into the CRISPR array over many days and to reconstruct the order in which spacers were delivered.  A constraint on storage capacity was dictated by a limit to total protospacer sequence.  From 15 recorded spacers, each with 27 bases and four bases per byte, the capacity was about 100 bytes.  Though the recording is distributed across the entire population and only partially encoded within any given cell, this method of information storage has intriguing potential.

\subsection{Mapping gene function and inheritance}

CRISPR-based methods have been employed to systematically analyze gene function.  CRISPRi was used to probe the interaction network of 300 essential genes in \emph{Bacillus subtilis} and to identify the contributions and relationships among genes involved in cell viability~\cite{107}.  Systematic knockdown of these genes confirmed the biological connection between genes of related processes, \emph{e.g.}, those responsible for cell wall biosynthesis and cell division, and revealed interesting connections between genes in distant functional groups, \emph{e.g.}, knockdown of a particular transcription gene resulted in cell wall defects.  The network of gene-gene connections that was established also uncovered genes involved in antibiotic resistance and cell morphology.  
	
A CRISPR-Cas9-based method has recently been developed to perform systematic genetic mapping~\cite{139}, which is the process of examining patterns of gene inheritance to identify chromosome location information, \emph{i.e.}, order and distances, for specific sequences that contribute to a particular phenotype.  Typical genetic mapping techniques rely on recombination events either during cellular meiosis or mitosis, however the recombination frequency is very low in both cases.  The CRISPR-based system developed by Sadhu and colleagues utilizes custom sgRNAs to generate a high density of mitotic recombination events in the \emph{Saccharomyces cerevisiae} (yeast) genome by introducing double-strand breaks at specific sites and facilitating repair by HDR.  This efficient method successfully identified DNA sequence differences that caused phenotypic variation.  It was able, for example, to find a single polymorphism that mapped to a sensitivity to manganese.

\subsection{Animal models}
\label{sec:AnimalModels}

CRISPR-Cas9 has aided the customizability of mammalian cell lines for specific needs and models~\cite{246}. Companies such as Addgene~\cite{H} and GenScript~\cite{G} have capitalized on CRISPR's versatility and specificity to generate stable cell lines with specified genomic deletions~\cite{246}, gene knockouts, or gene knock-ins~\cite{247}.  As \emph{in vitro} cell modifications became mastered, researchers turned to tackle \emph{in vivo} editing.  The first example of \emph{in vivo} CRISPR-Cas9-based genetic modification of endogenous genes was achieved in zebrafish embryos~\cite{34}.  Mouse models have been developed to study a variety of human ailments, including metabolic liver disease~\cite{249}, Huntington's disease~\cite{248}, and cancer chemotherapy~\cite{141}.

In the chemotherapy \emph{in vivo} mouse model, Braun and colleagues demonstrated the application of CRISPRi and CRISPRa to look at bone marrow treatment relapse~\cite{141}.  CRISPRi was used to inactivate the \emph{Trp53} gene, which transcribes the tumor protein p53 known to desensitize cells to a cytotoxic drug used in cancer chemotherapy, to model tumor cell resistance to therapy.  Additionally CRISPRa was compared with cDNA, a technique that makes DNA complementary to messenger RNAs in order to over-express the encoded protein of interest, to amplify a particular suicide enzyme gene that detoxifies DNA lesions.  Cell resistance to DNA damage via the chemotherapeutic agent temozolomide was significantly higher when CRISPRa was used.  A small library of sgRNAs was constructed to screen for genes that could delay tumor progression and increase therapeutic response.

\subsection{Human disease therapeutics}
\label{HumanDiseaseTherapeutics}

Researchers have started utilizing CRISPR-Cas9 as a gene therapy technique and were able to treat a gene mutation in dystrophin that causes Duchenne muscular dystrophy~\cite{108}.  They performed multiplex gene editing in human cells without significant toxicity to generate a large 336-kb deletion that had been previously established as a means to correct 62\% of these mutations.  Recent advances in CRISPR-Cas9-based editing in the human beta-globin gene have corrected a mutation in human embryos to reverse $\beta$-thalassemia~\cite{234} and in hematopoietic stem cells to cure sickle cell disease~\cite{235}.  While the side effects of germline editing in humans is still an open topic of research, \emph{ex vivo} modification of somatic cells is currently underway for lung, prostate, and renal cell cancer and HIV infection treatments~\cite{239}.

The first clinical trial of CRISPR-Cas9-modified T-cells given to humans was started in October 2016 with lung cancer patients, and more trials for \emph{in vivo} use in humans are underway for approval~\cite{E}.  Starting this year, the National Institutes of Health (NIH) plans to award \$190 million over six years to researchers committed to developing new somatic cell genome editors, delivery mechanisms, and assays for testing safety and efficacy for improved genome editing tools in patients~\cite{F}.  CRISPR-Cas9 has been at the forefront of current genome editing techniques and its continued improvement will no doubt be a priority.  To ensure safe and efficient editing systems, issues with \emph{in vivo} delivery of the CRISPR-Cas components~\cite{241} and the stability of Cas proteins in complex with the sgRNA~\cite{245} must be considered.  The human immune response is another factor that has recently been recognized, as the introduction of these components has been shown to elicit an innate response as well as the clonal expansion of Cas9-specific antibodies and T-cell receptors~\cite{243}.

\section{Conclusion and Suggestions for Future Work}
\label{sec:Conclusion}

In this review, we have outlined a wide range of experimental and theoretical work on the CRISPR-Cas system of prokaryotes.  The three mechanisms of adaptation, expression, and interference can be described as a Markov process, and they make use of a variety of CRISPR-specific proteins to protect the host cell.  Immunity against mobile genetic elements is achieved with spacer sequences chronicled in the CRISPR locus.  Modular sequence structures assist the crRNA:Cas protein complex in efficient and specific target recognition, and protein conformational changes regulate target cleavage.  HGT appears to have facilitated initial sharing of the CRISPR-Cas systems among diverse species, but there is selection against CRISPR in organisms that currently depend on HGT for pathogenicity.  More generally, CRISPR-Cas effectiveness is a determinant of loci evolution or elimination.  Population diversification of CRISPR loci rapidly occurs, since each strain adapts to combat its individual attackers.  Mathematical modeling has aided our understanding of the coevolutionary dynamics of CRISPR bacteria and phage.  CRISPR activity is regulated to minimize the cost associated with preparing for diverse threats and maximize energy efficiency.  Some particular species use their CRISPR systems for self-gene regulation and virulence, and additional unique uses will undoubtedly be discovered.  CRISPR-Cas provides a versatile platform for a range of gene editing, gene regulation, and imaging for biotechnology applications in bacteria, plants, and humans.

\subsection{Mechanisms of adaptive immunity}
\begin{itemize}
  \item What is the precise timescale of a bacterium's acquisition and utilization of CRISPR-Cas immunity, and how does this match up to the timescale of phage infection?
  \item Does a Markov model justly represent the CRISPR processes? Some experimental work has suggested more of an entanglement of mechanisms, and phenomena such as priming depend on past states of the system.
  \item It has been shown that the CRISPR locus has a maximum length of spacers, and spacer deletion occurs to allow new acquisitions.  Theoretical work could explore the plausibility of a bacteria cell having a dynamic maximum locus length, adaptable to different environmental situations. What would the relationship be between the maximum number of spacers in the locus and the diversity and evolution rate of phage in the environment?
\end{itemize}

\subsection{Evolution of CRISPR-Cas loci}
\begin{itemize}
  \item What are the principles that have governed the evolution of the highly diverse CRISPR types and subtypes in different species?
  \item In order for a CRISPR-Cas immune system to be effective, it must contain spacers that protect bacteria against phage that are specifically targeting them. What is the relative benefit of obtaining a whole CRISPR system with or without useful spacers through HGT versus already having a CRISPR system without useful spacers and needing to acquire new useful ones?  It is possible that HGT events have a lower probability, but are relevant on longer time scales.
  \item Phage are able to escape CRISPR-Cas recognition by mutating their protospacer. Above a certain threshold phage mutation rate, CRISPR is postulated to no longer be useful in bacteria.  If bacteria are in an environment with multiple phage types that have varying rates of mutation, are the bacteria more or less likely to have CRISPR?
\end{itemize}

\subsection{Stability and off-target activity}
\begin{itemize}
  \item Delivery method and disease background are important factors to consider when trying to implement CRISPR-based therapeutics in humans.  Can models of Cas protein immunogenicity determine an individual's immunological reaction and help to effectively design stable and deliverable CRISPR-Cas editing systems?
  \item More accurate modeling of the distribution of off-target effects is needed for biotechnology applications, especially in human cells.  What level of detail in mathematical modeling or computation simulations is necessary to predict sgRNA specificity?
    \item Currently Cas9 is the most popular CRISPR protein used in genomic engineering, due to its dual DNA binding and cleavage ability. Though the interference machinery of other CRISPR systems may be more difficult to harness, \emph{i.e.}, coordinating Cascade binding and Cas3 cleavage, they offer more specific control, as removal of one of the subunits can eliminate off-target binding.  How does the use of a modified Cascade affect the binding kinetics between the engineered sgRNA and target DNA sequence and the cleavage efficiency of Cas3?
\end{itemize}

\section*{Acknowledgement}
This work was partially supported by the Center for Theoretical Biological Physics at Rice University, Houston, TX 77005, USA.

\bibliographystyle{unsrt}
\bibliography{CRISPRreview}{}

\end{document}